\let\oldvec\vec
\documentclass[epj,nopacs]{svjour}
\let\vec\oldvec
\usepackage{amsfonts,amsmath,amssymb}
\usepackage{cite}
\usepackage{graphicx}
\usepackage[pdfstartview=FitH,colorlinks=true,linkcolor=blue,citecolor=blue,urlcolor=blue]{hyperref}
\usepackage{microtype}

\begin{document}

\title{Prethermalization in a quenched one-dimensional quantum fluid of light}
\subtitle{Intrinsic limits to the coherent propagation of a light beam in a nonlinear optical fiber}
\titlerunning{Prethermalization in a quenched one-dimensional quantum fluid of light}

\author{
Pierre-\'Elie Larr\'e%
\thanks{\email{pierre.larre@unitn.it}}
\and
Iacopo Carusotto%
\thanks{\email{carusott@science.unitn.it}}
}
\authorrunning{P.-\'E. Larr\'e and I. Carusotto}

\institute{INO-CNR BEC Center and Dipartimento di Fisica, Universit\`a di Trento, Via Sommarive 14, 38123 Povo, Italy}

\date{\today}

\abstract{We study the coherence properties of a laser beam after propagation along a one-dimensional lossless nonlinear optical waveguide. Within the paraxial, slowly-varying-envelope, and single-transverse-mode approximations, the quantum propagation of the light field in the nonlinear medium is mapped onto a quantum Gross--Pitaevskii-type evolution of a closed one-dimensional system of many interacting photons. Upon crossing the entrance and the back faces of the waveguide, the photon-photon interaction parameter undergoes two sudden jumps, resulting in a pair of quantum quenches of the system's Hamiltonian. In the weak-interaction regime, we use the modulus-phase Bogoliubov theory of dilute Bose gases to describe the quantum fluctuations of the fluid of light and predict that correlations typical of a prethermalized state emerge locally in their final form and propagate in a light-cone way at the Bogoliubov speed of sound in the photon fluid. This peculiar relaxation dynamics, visible in the light exiting the waveguide, results in a loss of long-lived coherence in the beam of light.}

\maketitle

\section{Introduction}
\label{Sec:Introduction}

In the presence of a significant Kerr optical nonlinearity, a many-photon light beam can behave as a quantum fluid of interacting bosons. This has opened the way to active experimental and theoretical investigations of many-body hydrodynamic and quantum features in photon-based systems, the research field of the so-called quantum fluids of light (see Ref.~\cite{Carusotto2013} for a review).

Numerous experimental studies have been performed in semiconductor planar microcavities, where the photon field strongly couples to the exciton one to form a mixed light-matter gas of interacting bosonic quasiparticles, the so-called exciton-polaritons \cite{Carusotto2013}. Among the most famous experimental investigations done on exciton-polariton fluids, one may cite the works demonstrating the occurrence of a Bose--Einstein-type condensation \cite{Kasprzak2006}, of a low-velocity superfluidlike flow around a localized material defect \cite{Amo2009}, of a Cherenkov radiation of Bogoliubov-type, linear, waves in a supersonic-flow regime, as well as the hydrodynamic nucleation of nonlinear excitations such as quantized vortices \cite{Nardin2011, Sanvitto2011} and dark solitons \cite{Amo2011, Grosso2011} past large impenetrable obstacles. For what concerns the corresponding theoretical literature, we invite the reader to consult review \cite{Carusotto2013}'s bibliography.

In semiconductor-planar-microcavity architectures, the dynamics of the light fluid is of driven-dissipative nature. This introduces a substantial complexity in the theoretical treatment of these systems and may be very detrimental for the experimental study of quantum phenomena.

An alternative, and promising, platform to investigate photon-fluid physics consists in the propagation of a quasimonochromatic electromagnetic wave through a nonabsorbing and cavityless, i.e., lossless, nonlinear optical medium of Kerr type. In contrast to what one has in cavity-based devices, the optical field in a cavityless, propagating, geometry obeys a fully conservative dynamics, usually described by the nonlinear Schr\"odinger equation of nonlinear optics within the paraxial and slowly-varying-envelope approximations \cite{Agrawal1995, Rosanov2002, Boyd2008, Larre2015a}.

At the mean-field level, this wave equation is analogous to the Gross--Pitaevskii equation of dilute atomic condensates \cite{Dalfovo1999, Pitaevskii2003} after having exchanged the roles played by the optical-axis coordinate $z$ and by the physical time parameter $t$: Light propagation in the $z$ direction is naturally described in terms of a Gross--Pitaevskii-type evolution of the photon field in a three-dimensional $(x,y,t)$ space where $t$ plays the role of the third spatial coordinate in addition to the transverse positions $x$ and $y$; the propagation constant of the wave and the dispersion parameter of the material provide two (\textit{a priori}) different effective masses to the light field in, respectively, the $(x,y)$ plane and the $t$ direction, the spatial profile of the medium's refractive index acts as an external potential, and finally, the Kerr-induced nonlinear shift of the medium's refractive index corresponds to contact interactions between photons. The initial conditions of the problem are fixed by the profile and the statistical properties of the light beam incident on the front face of the nonlinear dielectric and the final state may be reconstructed from the coherence and correlation patterns of the light emerging from the back, i.e., after the above-discussed Gross--Pitaevskii-type evolution across the medium.

This framework has been used in a number of experimental works devoted to the study of nonlinear features in propagating fluids of light, with a special attention dedicated to their relation to hydrodynamics and superfluidity aspects \cite{Vaupel1996, Wan2007, Jia2007, Wan2010a, Wan2010b, Jia2012, Vocke2015}. From the theoretical point of view, the nonlinear propagating geometry has also been subjected to numerous investigations, including, e.g., the study of superfluidlike behaviors in the flow of a photon fluid past a localized obstacle \cite{Pomeau1993, Hakim1997, Leboeuf2010, Carusotto2014, Larre2015b}, of nonlinear phenomena with light waves \cite{Dekel2007, Khamis2008, Dekel2009, Cohen2013}, and of the so-called acoustic Hawking radiation from analog black-hole horizons \cite{Fouxon2010, Fleurov2012, BarAd2013, Vinish2014}, the latter being accompanied with experimental works (see, e.g., Refs.~\cite{Elazar2012, Elazar2013}).

Building atop the pioneering theoretical works \cite{Lai1989a, Lai1989b} (consult also Ref.~\cite{Matsko2000}), a very general quantum theory of paraxial-light propagation in a bulk cavityless nonlinear optical medium was recently reported in Ref.~\cite{Larre2015a}. Within the above-discussed $z\longleftrightarrow t$ mapping, exact commutation relations for the quantum photon-field operator at equal propagation coordinates $z$ and different times $t$ were derived and the investigation of their practical consequences on the transmission of a beam of coherent light across a finite slab of weakly nonlinear dielectric immersed in air was carried out. Within the reformulation of light propagation in terms of an effective time evolution and as the Kerr nonlinearity is nonzero only inside the dielectric, the nonlinear propagating geometry was demonstrated to constitute a very simple realization of a pair of quantum quenches of the system's Hamiltonian in the photon-photon interaction parameter: The first one takes place at the entrance face of the nonlinear material and the second one occurs at its exit face. In the weak-nonlinearity regime, the main excitation process of the quenched quantum fluid of light was shown to consist in a sort of dynamical Casimir emission of correlated counterpropagating Bogoliubov-type waves, reflecting in peculiar features in the momentum distribution and in the near- and far-field two-body correlation functions of the transmitted beam of light. In this respect, the work done in Ref.~\cite{Larre2015a} made it possible to illustrate the power of the cavityless, propagating, geometry as a promising platform to investigate quantum dynamical features in closed, conservative, systems of many interacting bosons \cite{Kinoshita2006, Polkovnikov2011}.

In the present article, we pursue along these lines by investigating the consequences of the quantum quenches in the optical nonlinearity on the coherence of a beam of light propagating in a one-dimensional nonlinear optical waveguide. In such a configuration, the confining potential induced by the refractive-index difference between the core and the cladding of the waveguide in the transverse $x$, $y$ directions makes the dynamics unidimensional: The optical field now depends only on $z$ and $t$, with $z$ playing, as usual, the role of time and $t$ referencing spatial positions along the one-dimensional quantum fluid of light. We calculate the degree of first-order coherence of the transmitted beam of light in the weak-nonlinearity regime and predict that correlations typical of the installation of a prethermalized state \cite{Berges2004} emerge locally in their eventual form and propagate in a light-cone way at the Bogoliubov speed of sound in the nonlinear medium, in a similar way as it was recently observed in quantum-quenched phase-fluctuating one-dimensional atomic Bose gases \cite{Gring2012, Kuhnert2013, Langen2013}. This results in a loss of long-lived coherence in the transmitted beam of light, which could have detrimental practical consequences, e.g., in fiber-optic communication.

The paper is organized as follows. First, in Sect.~\ref{Sec:ClassicalWaveEquation}, we review the paraxial propagation of a quasimonochromatic light wave through a one-dimensional (1D) optical waveguide presenting a Kerr nonlinearity, starting from the full three-dimensional (3D), bulk, paraxial-optics problem. In the single-transverse-mode regime, the transverse profile of the light beam is frozen in the ground state, described by a two-dimensional (2D) Schr\"odinger-type equation, while the longitudinal motion is ruled by a 1D nonlinear wave equation similar to the time-dependent Gross--Pitaevskii equation of quasi-1D ultracold dilute Bose gases. Then, in Sect.~\ref{Sec:QuantumTheory}, we carry out the quantization of the classical 1D propagation equation, making use of the general 3D quantum field theory investigated in Ref.~\cite{Larre2015a}. The resulting 1D quantum theory makes it possible to describe quantum features in the 1D light fluid for generic values of the photon-photon interaction parameter. We also discuss how the quantum fluctuations of the 1D optical field in the dilute-gas limit can be treated within the modulus-phase Bogoliubov theory of weakly interacting Bose gases. Assuming that the photon fluid propagating in the waveguide is well in the dilute regime, we calculate in Sect.~\ref{Sec:LightCoherenceInResponseToQuantumQuenchesInTheKerrNonlinearity} the degree of first-order coherence of the transmitted beam of light, after propagation along the waveguide. The features that it displays are interpreted in terms of a dynamical local emergence of a prethermalized state in a quenched 1D quantum system of many weakly interacting photons. Finally, we draw in Sect.~\ref{Sec:Conclusion} our conclusions and give prospects to the present work.

\section{Classical wave equation}
\label{Sec:ClassicalWaveEquation}

In this section, we review the classical equation of motion of a coherent quasimonochromatic light beam propagating along a Kerr-type 1D optical waveguide. As reviewed in Sect.~\ref{SubSubSec:OneDimensionalReduction}, our approach for describing the 1D motion of the optical field is based on a 1D reduction of the 3D nonlinear wave equation exposed in Sect.~\ref{SubSec:ThreeDimensionalPropagationEquation}, in a way analog to what it is standardly done in the theory of atomic Bose gases to describe the motion of 1D-confined atom bosons. The formal analogy existing between the 1D propagation equation and the time-dependent Gross--Pitaevskii equation of quasi-1D dilute Bose--Einstein condensates will lead us in Sect.~\ref{SubSubSec:GrossPitaevskiiTypeTimeEvolution} to reformulate the longitudinal motion of the photon field in terms of an effective temporal evolution in a 1D space spanned by the physical time parameter.

\subsection{Three-dimensional propagation equation}
\label{SubSec:ThreeDimensionalPropagationEquation}

We consider the propagation of a laser beam in a nonabsorbing 1D optical waveguide of length $L$ along the $z$ axis (the propagation occurs in the positive-$z$ direction and the coordinate origin corresponds to the center of the entrance face of the waveguide) and of total refractive index
\begin{equation}
\label{Eq:RefractiveIndex}
n(\mathbf{x},\omega)=n(\omega)+\mathrm{\Delta}n(\mathbf{x},\omega)+n_{2}(\omega)\,|\mathcal{E}|^{2}.
\end{equation}
In this equation, the homogeneous contribution $n(\omega)$, function of the optical angular frequency $\omega$, takes into account the chromatic dispersion of the waveguide, the linear shift $\mathrm{\Delta}n(\mathbf{x},\omega)$, function of $\mathbf{x}=(x,y)$, describes the transverse spatial profile of the total refractive index, and the nonlinear shift $n_{2}(\omega)\,|\mathcal{E}|^{2}$ originates from the optical nonlinearity of the waveguide, of Kerr type. The latter is proportional to the square modulus of the complex amplitude, the envelope, $\mathcal{E}$ of the laser wave's electric field
\begin{equation}
\label{Eq:ElectricField}
E(\mathbf{x},z,t)=\tfrac{1}{2}\,\mathcal{E}(\mathbf{x},z,t)\,\mathrm{e}^{\mathrm{i}(\beta_{0}z-\omega_{0}t)}+\mathrm{c.c.}
\end{equation}
(``c.c.'' stands for ``complex conjugate'') of central angular frequency $\omega_{0}$ and propagation constant $\beta_{0}=\beta(\omega_{0})$ in the positive-$z$ direction, where, denoting $c_{0}$ the vacuum speed of light, $\beta(\omega)=n(\omega)\,\omega/c_{0}$. Obviously, the total refractive index \eqref{Eq:RefractiveIndex} has to be transversally higher in the core of the waveguide than at its surface to make an optical confinement possible, that is, to realize a guiding configuration. Finally, the optical medium constituting the waveguide is supposed to be nonmagnetic and devoid of free electric charges and the laser wave is assumed to maintain its polarization in the course of its propagation in the waveguide so that a scalar appoach is valid, as implicitly considered in Eq.~\eqref{Eq:ElectricField}.

Making use of the usual paraxial and slowly-varying-envelope approximations (see, e.g., Refs.~\cite{Agrawal1995, Rosanov2002, Boyd2008}), Maxwell's equations supplemented by Eqs.~\eqref{Eq:RefractiveIndex} and \eqref{Eq:ElectricField} lead to the following 3D nonlinear propagation equation for the electric field's envelope $\mathcal{E}(\mathbf{x},z,t)$ (see Ref.~\cite{Larre2015a} and references therein):
\begin{align}
\notag
\mathrm{i}\,\frac{\partial\mathcal{E}}{\partial z}=&\left.-\frac{1}{2\,\beta_{0}}\,\nabla^{2}\mathcal{E}+\frac{D_{0}}{2}\,\frac{\partial^{2}\mathcal{E}}{\partial t^{2}}-\frac{\mathrm{i}}{v_{0}}\,\frac{\partial\mathcal{E}}{\partial t}\right. \\
\label{Eq:3DGPE}
&\left.+\,U(\mathbf{x})\,\mathcal{E}+g_{\mathrm{3D}}\,|\mathcal{E}|^{2}\,\mathcal{E}.\right.
\end{align}
In this equation, $\nabla=(\partial_{x},\partial_{y})$ denotes the nabla operator in the $\mathbf{x}=(x,y)$ plane and the constants $v_{0}=v(\omega_{0})$ and $D_{0}=D(\omega_{0})$ are the group velocity $v(\omega)=[\mathrm{d}\beta(\omega)/\mathrm{d}\omega]^{-1}$ of the photons in the medium and the group-velocity dispersion $D(\omega)=\mathrm{d}^{2}\beta(\omega)/\mathrm{d}\omega^{2}$ of the waveguide's material evaluated at the carrier's angular frequency $\omega_{0}$; finally, the parameters $U(\mathbf{x})$ and $g_{\mathrm{3D}}$ are defined as
\begin{equation}
\label{Eq:ExternalPotential3DInteractionConstant}
U(\mathbf{x})=-k_{0}\,\mathrm{\Delta}n(\mathbf{x},\omega_{0})\quad\text{and}\quad g_{\mathrm{3D}}=-k_{0}\,n_{2}(\omega_{0}),
\end{equation}
where $k_{0}=\omega_{0}/c_{0}=\beta_{0}/n_{0}$ [with $n_{0}=n(\omega_{0})$] is the propagation constant of the laser beam in air, i.e., outside the waveguide, in the region where $z<0$ or $z>L$.

\subsection{One-dimensional propagation equation}
\label{SubSec:OneDimensionalPropagationEquation}

\subsubsection{One-dimensional reduction}
\label{SubSubSec:OneDimensionalReduction}

Within the adiabatic approximation (see, e.g., Refs.~\cite{Olshanii1998, Jackson1998, Leboeuf2001, Pavloff2002} specific to the physics of Bose--Einstein condensation), the transverse [that is, in the $(x,y)$ plane] light profile is not affected by optical powers at points other than $z$ and has a weak dependence on both $z$ and $t$. In this case, one may expand the envelope $\mathcal{E}$ of the electric field as
\begin{equation}
\label{Eq:AdiabaticApproximation}
\mathcal{E}(\mathbf{x},z,t)=\mathrm{\Psi}(z,t)\,\mathrm{\Phi}(\mathbf{x};|\mathrm{\Psi}(z,t)|^{2}),
\end{equation}
where $\mathrm{\Psi}$ is the function describing the longitudinal motion and $\mathrm{\Phi}$, normalized to unity, $\int\mathrm{d}^{2}x\;|\mathrm{\Phi}|^{2}=1$ ($\mathrm{d}^{2}x=\mathrm{d}x\,\mathrm{d}y$), is the function describing the transverse one; in the adiabatic approximation, the latter depends parametrically on the square modulus of $\mathrm{\Psi}$,
\begin{equation}
\label{Eq:Power}
|\mathrm{\Psi}(z,t)|^{2}=\int\mathrm{d}^{2}x\;|\mathcal{E}(\mathbf{x},z,t)|^{2},
\end{equation}
which actually fixes the electromagnetic power\footnote{The intensity of the propagating quasimonochromatic wave of central angular frequency $\omega_{0}$ is expressed as a function of the square modulus of the complex amplitude $\mathcal{E}$ of its electric field as $\mathcal{I}=\frac{1}{2}\,c_{0}\,\varepsilon_{0}\,n_{0}\,|\mathcal{E}|^{2}$, where $\varepsilon_{0}$ is the vacuum permittivity. As a result, the power $\mathcal{P}=\int\mathrm{d}^{2}x\;\mathcal{I}$ carried by this wave may be deduced from $|\mathrm{\Psi}|^{2}$ as $\mathcal{P}=\frac{1}{2}\,c_{0}\,\varepsilon_{0}\,n_{0}\,|\mathrm{\Psi}|^{2}$.} propagating in the waveguide. Inserting Eq.~\eqref{Eq:AdiabaticApproximation} into the 3D wave equation \eqref{Eq:3DGPE}, one obtains, at leading order in the adiabatic approximation,
\begin{equation}
\label{Eq:1DWaveEquation}
\mathrm{i}\,\frac{\partial\mathrm{\Psi}}{\partial z}=\frac{D_{0}}{2}\,\frac{\partial^{2}\mathrm{\Psi}}{\partial t^{2}}-\frac{\mathrm{i}}{v_{0}}\,\frac{\partial\mathrm{\Psi}}{\partial t}+\kappa(|\mathrm{\Psi}|^{2})\,\mathrm{\Psi},
\end{equation}
where the nonlinear term $\kappa(|\mathrm{\Psi}|^{2})$ is determined as a function of $|\mathrm{\Psi}|^{2}$ from the equation of motion of the transverse profile,
\begin{equation}
\label{Eq:Nonlinearity}
\kappa(|\mathrm{\Psi}|^{2})\,\mathrm{\Phi}=\bigg[{-}\frac{1}{2\,\beta_{0}}\,\nabla^{2}+U(\mathbf{x})+g_{\mathrm{3D}}\,|\mathrm{\Psi}|^{2}\,|\mathrm{\Phi}|^{2}\bigg]\,\mathrm{\Phi}.
\end{equation}

When the term proportional to $g_{\mathrm{3D}}$ in the right-hand side of Eq.~\eqref{Eq:Nonlinearity} is small, a perturbative solution of Eq.~\eqref{Eq:Nonlinearity} yields $\kappa(|\mathrm{\Psi}|^{2})=\kappa_{0}+g_{\mathrm{1D}}\,|\mathrm{\Psi}|^{2}$, where $\kappa_{0}$ is the eigenvalue associated to the ground state $\mathrm{\Phi}_{0}(\mathbf{x})$ of the Schr\"odinger-type operator $-\nabla^{2}/(2\,\beta_{0})+U(\mathbf{x})$ \{$\kappa_{0}\,\mathrm{\Phi}_{0}=[-\nabla^{2}/(2\,\beta_{0})+U(\mathbf{x})]\,\mathrm{\Phi}_{0}$\}, i.e., to the fundamental transverse modal function of the waveguide, and where
\begin{equation}
\label{Eq:1DInteractionConstant}
g_{\mathrm{1D}}=\frac{g_{\mathrm{3D}}}{A_{\mathrm{eff}}},
\end{equation}
\begin{equation}
\label{Eq:EffectiveModeArea}
A_{\mathrm{eff}}=\frac{(\int\mathrm{d}^{2}x\;|\mathrm{\Phi}_{0}|^{2})^{2}}{\int\mathrm{d}^{2}x\;|\mathrm{\Phi}_{0}|^{4}}=\frac{1}{\int\mathrm{d}^{2}x\;|\mathrm{\Phi}_{0}|^{4}}
\end{equation}
denoting the effective area of the fundamental transverse mode of the waveguide \cite{Agrawal1995}. In this case, the 1D nonlinear wave equation \eqref{Eq:1DWaveEquation} becomes
\begin{equation}
\label{Eq:1DGPE}
\mathrm{i}\,\frac{\partial\mathrm{\Psi}}{\partial z}=\frac{D_{0}}{2}\,\frac{\partial^{2}\mathrm{\Psi}}{\partial t^{2}}-\frac{\mathrm{i}}{v_{0}}\,\frac{\partial\mathrm{\Psi}}{\partial t}+\kappa_{0}\,\mathrm{\Psi}+g_{\mathrm{1D}}\,|\mathrm{\Psi}|^{2}\,\mathrm{\Psi}.
\end{equation}

In view of experiments, it is important to explicit the domain of validity of Eq.~\eqref{Eq:1DGPE}. According to what precedes, the perturbative expansion $\kappa(|\mathrm{\Psi}|^{2})=\kappa_{0}+g_{\mathrm{1D}}\,|\mathrm{\Psi}|^{2}$, and so the resulting 1D nonlinear wave equation \eqref{Eq:1DGPE}, is accurate provided that the 1D nonlinear term $g_{\mathrm{1D}}\,|\mathrm{\Psi}|^{2}$ is a weak correction to the ground-state eigenvalue $\kappa_{0}$ of the radial operator $-\nabla^{2}/(2\,\beta_{0})+U(\mathbf{x})$,
\begin{equation}
\label{Eq:SingleTransverseModeCondition}
|g_{\mathrm{1D}}|\,|\mathrm{\Psi}|^{2}\ll|\kappa_{0}|.
\end{equation}
This condition ensures that all radial motion is reduced to zero-point oscillations, i.e., that the waveguide is characterized by one, and just one, transverse mode: the fundamental one of wavefunction $\mathrm{\Phi}_{0}(\mathbf{x})$. In the present optical context, the constraint \eqref{Eq:SingleTransverseModeCondition} is the analog of the single-transverse-mode condition \cite{Pitaevskii2003, Menotti2002}
\begin{equation}
\label{Eq:SingleTransverseModeConditionBEC}
|\lambda_{\mathrm{1D}}|\,n_{\mathrm{1D}}\ll\hbar\,\omega_{\perp}
\end{equation}
in the theory of quasi-1D ultracold Bose gases, $\lambda_{\mathrm{1D}}$, $n_{\mathrm{1D}}$, and $\hbar\,\omega_{\perp}$ denoting the effective 1D atom-atom interaction constant, the 1D density, and the energy of the transverse ground state, respectively, of the atomic Bose gas. Taking advantage of the definitions \eqref{Eq:ExternalPotential3DInteractionConstant} and \eqref{Eq:1DInteractionConstant}, the inequality \eqref{Eq:SingleTransverseModeCondition} may be reformulated in terms of the optical constants of the problem as
\begin{equation}
\label{Eq:SingleTransverseModeConditionExperiments}
|\mathrm{\Delta}n_{\mathrm{NL}}|\ll\frac{|\kappa_{0}|}{k_{0}},
\end{equation}
where
\begin{equation}
\label{Eq:NonlinearRefractiveIndex}
\mathrm{\Delta}n_{\mathrm{NL}}=\frac{\tilde{n}_{2}(\omega_{0})\,\mathcal{P}}{A_{\mathrm{eff}}}
\end{equation}
is the nonlinear shift of the waveguide's refractive index \eqref{Eq:RefractiveIndex} expressed in terms of the Kerr-nonlinearity coefficient $\tilde{n}_{2}(\omega_{0})=2\,n_{2}(\omega_{0})/(c_{0}\,\varepsilon_{0}\,n_{0})$ in intensity units (that is, in $\mathrm{m}^{2}/\mathrm{W}$ in SI units), of the net power $\mathcal{P}=\tfrac{1}{2}\,c_{0}\,\varepsilon_{0}\,n_{0}\,|\mathrm{\Psi}|^{2}$ carried by the quasimonochromatic wave \eqref{Eq:ElectricField}, and of the waveguide's effective transverse-mode area $A_{\mathrm{eff}}$.

As one may note, the shape of the ``confining potential'' $U(\mathbf{x})=-k_{0}\,\mathrm{\Delta}n(\mathbf{x},\omega_{0})$ has never been explicitly specified, as a result of which the equations obtained up to now are very general. However, always in view of experiments, it is important to provide explicit formulas for standard optical confinements, e.g., of parabolic or square-well shape. This is what we do in the Appendix.

\subsubsection{Gross--Pitaevskii-type time evolution}
\label{SubSubSec:GrossPitaevskiiTypeTimeEvolution}

The formal analogy there exists between the 1D nonlinear propagation equation \eqref{Eq:1DGPE} and the time-dependent Gross--Pitaevskii equation of quasi-1D dilute Bose--Einstein condensates \cite{Jackson1998, Leboeuf2001, Pavloff2002, Pitaevskii2003} is straightforward. The longitudinal photon field $\mathrm{\Psi}$ (its square modulus $|\mathrm{\Psi}|^{2}$, respectively) plays the role of the 1D order parameter of the condensate (of its density, respectively), the optical-axis, propagation, coordinate $z$ that of time, the time parameter $t$ corresponds to the coordinate referencing points along the 1D condensate, the group-velocity dispersion $D_{0}$, or more precisely $-1/D_{0}$, acts as a mass, $\kappa_{0}$ plays the role of the energy of the transverse ground state, and $g_{\mathrm{1D}}$ corresponds to the effective 1D atom-atom interaction constant. This correspondence makes it possible to reformulate the propagation of the wave \eqref{Eq:ElectricField} along the optical waveguide \eqref{Eq:RefractiveIndex} in the language of quantum hydrodynamics. Mention also that the rigid-drift term $-\mathrm{i}\,\partial_{t}\mathrm{\Psi}/v_{0}$ in the right-hand side of the 1D equation \eqref{Eq:1DGPE} originates from the fact that the photons propagate in the waveguide at the group velocity $v_{0}$.

To reexpress the propagation of the photon field $\mathrm{\Psi}(z,t)$ in the positive-$z$ direction in terms of a temporal evolution in a 1D space spanned by the physical time parameter $t$, we introduce the following coordinates:
\begin{equation}
\label{Eq:GPECoordinates}
\tau=\frac{z}{v_{0}}\quad\text{and}\quad\zeta=v_{0}\,t-z,
\end{equation}
homogeneous to a time and a length, respectively. In these new variables, Eq.~\eqref{Eq:1DGPE} multiplied by the reduced Planck constant $\hbar$ and the group velocity $v_{0}$ reads
\begin{equation}
\label{Eq:ClassicalGPE}
\mathrm{i}\,\hbar\,\frac{\partial\mathrm{\Psi}}{\partial\tau}=-\frac{\hbar^{2}}{2\,m}\,\frac{\partial^{2}\mathrm{\Psi}}{\partial\zeta^{2}}+E_{0}\,\mathrm{\Psi}+g\,|\mathrm{\Psi}|^{2}\,\mathrm{\Psi},
\end{equation}
where the longitudinal optical field $\mathrm{\Psi}$ now has to be considered as a function of $\zeta$ and $\tau$. We end up with an effective time-dependent 1D Gross--Pitaevskii equation for the ``order parameter'' $\mathrm{\Psi}(\zeta,\tau)$ of the 1D ``quantum fluid of light'' where
\begin{equation}
\label{Eq:Mass}
m=-\frac{\hbar}{v_{0}^{3}\,D_{0}^{\vphantom{3}}}
\end{equation}
is the photon effective mass, function of the group-velocity dispersion $D(\omega)$ of the waveguide at $\omega=\omega_{0}$, $E_{0}=\hbar\,v_{0}\,\kappa_{0}$, and
\begin{equation}
\label{Eq:InteractionConstant}
g=\hbar\,v_{0}\,g_{\mathrm{1D}}
\end{equation}
is the photon-photon (contactlike) interaction constant.

According to Eq.~\eqref{Eq:Mass}, we get a positive (negative) mass $m$ when $D_{0}<0$ ($D_{0}>0$), i.e., when the waveguide is characterized by an anomalous (normal) group-velocity dispersion at $\omega_{0}$. On the other hand, according to Eqs.~\eqref{Eq:ExternalPotential3DInteractionConstant}, \eqref{Eq:1DInteractionConstant}, and \eqref{Eq:InteractionConstant}, one has repulsive interactions between photons ($g>0$) when the medium is self-defocusing [$n_{2}(\omega_{0})<0$] and attractive interactions ($g<0$) when it is self-focusing [$n_{2}(\omega_{0})>0$]. In Sect.~\ref{SubSec:ModulusPhaseBogoliubovTheoryOfQuantumFluctuations}, we shall see that the dynamical stability of the fluid of light requires that both $m$ and $g$, i.e., both $D_{0}$ and $n_{2}(\omega_{0})$, have the same sign.

Finally, note that the second of Eqs.~\eqref{Eq:GPECoordinates} corresponds to the relationship linking the coordinate systems $\{\mathbf{x},z,t\}$ and $\{\mathbf{x},\zeta,t\}$ of two Galilean frames of reference of relative velocity $v_{0}$ in the $z$ direction. Thus, it is natural that the first-order derivative with respect to $t$ in the right-hand side of Eq.~\eqref{Eq:1DGPE} is no longer present in Eq.~\eqref{Eq:ClassicalGPE}.

\section{Quantum theory}
\label{Sec:QuantumTheory}

In order to describe nonclassical---i.e., quantum---features in the paraxial beam of light propagating in the 1D nonlinear optical waveguide considered in Sect.~\ref{Sec:ClassicalWaveEquation}, the classical longitudinal light field $\mathrm{\Psi}(\zeta,\tau)$ obeying Eq.~\eqref{Eq:ClassicalGPE} has to be replaced with a quantum field operator satisfying suitable equal-time-$\tau$, that is, within the $z\longleftrightarrow t$ mapping introduced in Sect.~\ref{SubSubSec:GrossPitaevskiiTypeTimeEvolution}, equal-$z$, boson commutation relations, as done in the pioneering references \cite{Lai1989a, Lai1989b} to investigate quantum soliton propagation in optical fibers. In Sect.~\ref{SubSec:Quantization}, we will demonstrate that this may be fully achieved on the basis of the general 3D quantum theory studied in Ref.~\cite{Larre2015a}. Then, in Sect.~\ref{SubSec:ModulusPhaseBogoliubovTheoryOfQuantumFluctuations}, considering that the 1D quantum fluid of light is well within the weakly interacting regime, we shall apply the modulus-phase Bogoliubov theory of dilute atomic Bose gases to describe its quantum fluctuations.

\subsection{Quantization}
\label{SubSec:Quantization}

To quantize the classical Gross--Pitaevskii equation \eqref{Eq:ClassicalGPE}, one can start from the quantum field theory associated to the 3D nonlinear wave equation \eqref{Eq:3DGPE}. It was shown in Ref.~\cite{Larre2015a} that the quantum field operator $\hat{\mathcal{E}}(\mathbf{x},\zeta,\tau)$ corresponding to the solution of the classical equation \eqref{Eq:3DGPE} rewritten in the coordinates \eqref{Eq:GPECoordinates} has to satisfy the equal-$\tau$ [that is, equal-$z$; see the first of Eqs.~\eqref{Eq:GPECoordinates}] boson commutation relations
\begin{subequations}
\label{Eq:3DCommutationRelations}
\begin{align}
\notag
[\hat{\mathcal{E}}(\mathbf{x},\zeta,\tau),\hat{\mathcal{E}}^{\dag}(\mathbf{x}',\zeta',\tau)]&\left.=\mathcal{N}\,\frac{\hbar\,\omega_{0}}{\varepsilon_{0}}\right. \\
\label{Eq:3DCommutationRelations1}
&\left.\hphantom{=}\times\delta^{(2)}(\mathbf{x}-\mathbf{x}')\,\delta(\zeta-\zeta'),\right. \\
\label{Eq:3DCommutationRelations2}
[\hat{\mathcal{E}}(\mathbf{x},\zeta,\tau),\hat{\mathcal{E}}(\mathbf{x}',\zeta',\tau)]&\left.=0,\right.
\end{align}
\end{subequations}
where
\begin{equation}
\label{Eq:NormalizationConstant}
\mathcal{N}=\frac{2\,v_{0}}{n_{0}\,c_{0}}
\end{equation}
is a normalization parameter whose expression in terms of the optical constants $v_{0}$, $n_{0}$, and $c_{0}$ may be deduced from \textit{ab initio} classical-electrodynamics calculations\footnote{See Ref.~\cite{Larre2015a}. Note that a factor $1/2$ is missing in Eq.~(24) of this reference, the mistake coming from a double counting of the dynamical variables involved in the classical field theory investigated in Sect.~III B.}.

The quantization of the 1D classical theory is straightforwardly accomplished by replacing the classical optical field $\mathrm{\Psi}(\zeta,\tau)$ with the projection
\begin{equation}
\label{Eq:QuantumField}
\hat{\mathrm{\Psi}}(\zeta,\tau)=\int\mathrm{d}^{2}x\;\mathrm{\Phi}_{0}^{\ast}(\mathbf{x})\,\hat{\mathcal{E}}(\mathbf{x},\zeta,\tau)
\end{equation}
of the total 3D quantum field operator $\hat{\mathcal{E}}(\mathbf{x},\zeta,\tau)$ onto the transverse ground state of wavefunction $\mathrm{\Phi}_{0}(\mathbf{x})$, for instance given (see the Appendix) by the second of Eqs.~\eqref{Eq:GroundStateHarmonicConfinement} in the case of a parabolic confining potential or by Eq.~\eqref{Eq:GroundStateSquareWellConfinement2} in the case of a square-well-shaped one. Multiplying Eq.~\eqref{Eq:3DCommutationRelations1} [Eq.~\eqref{Eq:3DCommutationRelations2}] by $\mathrm{\Phi}_{0}^{\ast}(\mathbf{x})\,\mathrm{\Phi}_{0}^{\vphantom{\ast}}(\mathbf{x}')$ [by $\mathrm{\Phi}_{0}^{\ast}(\mathbf{x})\,\mathrm{\Phi}_{0}^{\ast}(\mathbf{x}')$], integrating over both $\mathbf{x}$ and $\mathbf{x}'$, and using the normalization condition $\int\mathrm{d}^{2}x\;|\mathrm{\Phi}_{0}|^{2}=1$, one eventually finds that \eqref{Eq:QuantumField} obeys the same-$\tau$ commutation rules
\begin{subequations}
\label{Eq:1DCommutationRelations}
\begin{align}
\label{Eq:1DCommutationRelations1}
[\hat{\mathrm{\Psi}}(\zeta,\tau),\hat{\mathrm{\Psi}}^{\dag}(\zeta',\tau)]&=\mathcal{N}\,\frac{\hbar\,\omega_{0}}{\varepsilon_{0}}\,\delta(\zeta-\zeta'), \\
\label{Eq:1DCommutationRelations2}
[\hat{\mathrm{\Psi}}(\zeta,\tau),\hat{\mathrm{\Psi}}(\zeta',\tau)]&=0.
\end{align}
\end{subequations}

The normally-ordered quantized version of the classical 1D equation \eqref{Eq:ClassicalGPE} reads
\begin{equation}
\label{Eq:QuantumGPE}
\mathrm{i}\,\hbar\,\frac{\partial\hat{\mathrm{\Psi}}}{\partial\tau}=-\frac{\hbar^{2}}{2\,m}\,\frac{\partial^{2}\hat{\mathrm{\Psi}}}{\partial\zeta^{2}}+E_{0}\,\hat{\mathrm{\Psi}}+g\,\hat{\mathrm{\Psi}}^{\dag}\,\hat{\mathrm{\Psi}}\,\hat{\mathrm{\Psi}}.
\end{equation}
As originally pointed out in the pioneering works \cite{Lai1989a, Lai1989b}, it has the form of a quantum nonlinear Schr\"odinger equation, i.e., of the quantum Gross--Pitaevskii equation in the context of atomic Bose gases \cite{Pitaevskii2003}. It may be rewritten in the form of a quantum mechanical evolution equation in the Heisenberg representation, i.e., in the form
\begin{equation}
\label{Eq:HeisenbergEquationOfMotion}
\mathrm{i}\,\hbar\,\frac{\partial\hat{\mathrm{\Psi}}}{\partial\tau}=[\hat{\mathrm{\Psi}},\hat{H}(\tau)],
\end{equation}
where
\begin{align}
\notag
\hat{H}(\tau)=&\left.\Big(\mathcal{N}\,\frac{\hbar\,\omega_{0}}{\varepsilon_{0}}\Big)^{-1}\right. \\
\label{Eq:Hamiltonian}
&\left.\times\int\mathrm{d}\zeta\;\hat{\mathrm{\Psi}}^{\dag}\,\bigg({-}\frac{\hbar^{2}}{2\,m}\,\frac{\partial^{2}}{\partial\zeta^{2}}+E_{0}+\frac{g}{2}\,\hat{\mathrm{\Psi}}^{\dag}\,\hat{\mathrm{\Psi}}\bigg)\,\hat{\mathrm{\Psi}}\right.
\end{align}
is the normally ordered many-body quantum Hamiltonian operator of the optical system. The first contribution in the integral over $\zeta$ is the kinetic term in the $\zeta=v_{0}\,t-z$ direction at a given position $z$ along the optical axis, that is, at a given time $\tau=z/v_{0}$, the second one corresponds to the transverse-ground-state energy shift due to the optical confinement in the $x$, $y$ directions, and the last one accounts for the two-photon interactions mediated by the Kerr nonlinearity of the dielectric constituting the optical waveguide. Note that referring to $\hat{H}(\tau)$ as a ``Hamiltonian'' follows from its interpretation as an evolution operator, although it physically corresponds to the $z$ component of the momentum operator, as it generates translation in the $z$ direction.

The quantum Gross--Pitaevskii equation \eqref{Eq:QuantumGPE} and the boson commutation relations \eqref{Eq:1DCommutationRelations} at equal times $\tau$ (i.e., at equal propagation distances $z$) and at different positions $\zeta$ (i.e., at different physical times $t$) constitute the heart of the 1D quantum field theory investigated in this work. According to Sect.~\ref{SubSubSec:OneDimensionalReduction}, the classical counterpart \eqref{Eq:ClassicalGPE} of the quantum equation \eqref{Eq:QuantumGPE} is valid as long as the single-transverse-mode condition \eqref{Eq:SingleTransverseModeConditionExperiments} is fulfilled. This naturally transfers to Eq.~\eqref{Eq:QuantumGPE} upon the quantization procedure \eqref{Eq:QuantumField}. At this stage of the paper, no hypothesis has been made on the strength of the two-photon interactions, as a result of which Eq.~\eqref{Eq:QuantumGPE} makes it possible to describe quantum features in the 1D fluid of light for generic values of the photon-photon interaction parameter.

In the following, we nevertheless shall focus our attention on the weak-interaction regime, that is, the dilute or superfluid limit. This latter may be obtained by requiring that the interaction term $g\,\hat{\mathrm{\Psi}}^{\dag}\,\hat{\mathrm{\Psi}}\,\hat{\mathrm{\Psi}}$ in the 1D equation \eqref{Eq:QuantumGPE} is small compared to the kinetic term $-\hbar^{2}\,\partial_{\zeta\zeta}\hat{\mathrm{\Psi}}/(2\,m)=-\hbar^{2}\,\partial_{tt}\hat{\mathrm{\Psi}}/(2\,v_{0}^{2}\,m)$. Approximating $\hat{\mathrm{\Psi}}^{\dag}\,\hat{\mathrm{\Psi}}$ by $|\mathrm{\Psi}|^{2}$ and evaluating the second-order derivative $\partial_{tt}\hat{\mathrm{\Psi}}$ as $\partial_{tt}\hat{\mathrm{\Psi}}\sim\mathcal{F}^{2}\,\hat{\mathrm{\Psi}}$, where $\mathcal{F}=\mathcal{P}/(\hbar\,\omega_{0})=\frac{1}{2}\,c_{0}\,\varepsilon_{0}\,n_{0}\,|\mathrm{\Psi}|^{2}/(\hbar\,\omega_{0})$, which is homogeneous to the inverse of a time, denotes the photon flux (also called photon power in photon-counting physics), the dilute-gas regime is reached when
\begin{equation}
\label{Eq:DiluteGasCondition}
|g|\,|\mathrm{\Psi}|^{2}\ll\frac{\hbar^{2}}{2\,v_{0}^{2}\,|m|}\,\mathcal{F}^{2}\propto\frac{\hbar^{2}}{2\,|m|}\,(|\mathrm{\Psi}|^{2})^{2},
\end{equation}
which corresponds to the weak-interaction limit \cite{Pitaevskii2003, Menotti2002}
\begin{equation}
\label{Eq:DiluteGasConditionBEC}
|\lambda_{\mathrm{1D}}^{\vphantom{2}}|\,n_{\mathrm{1D}}^{\vphantom{2}}\ll\frac{\hbar^{2}}{2\,M}\,n_{\mathrm{1D}}^{2}
\end{equation}
for a 1D atomic fluid, where, as introduced in Sect.~\ref{SubSubSec:OneDimensionalReduction}, $\lambda_{\mathrm{1D}}$ and $n_{\mathrm{1D}}$ respectively denote the 1D atom-atom interaction constant and the 1D atom density, and $M$ is the atom mass. Using Eqs.~\eqref{Eq:ExternalPotential3DInteractionConstant}, \eqref{Eq:1DInteractionConstant}, \eqref{Eq:Mass}, and \eqref{Eq:InteractionConstant}, eliminating the beam's power $\mathcal{P}$ from the left-hand side of \eqref{Eq:DiluteGasCondition}, and multiplying the inequality by $|n_{2}(\omega_{0})|/A_{\mathrm{eff}}$, the constraint \eqref{Eq:DiluteGasCondition} may be reformulated as the following condition for the nonlinear shift $\mathrm{\Delta}n_{\mathrm{NL}}$ of the waveguide's refractive index defined in Eq.~\eqref{Eq:NonlinearRefractiveIndex}:
\begin{equation}
\label{Eq:DiluteGasConditionExperiments}
\frac{2\,(\hbar\,\omega_{0})^{2}\,k_{0}\,|\tilde{n}_{2}(\omega_{0})|^{2}}{|D_{0}|\,A_{\mathrm{eff}}^{2}}\ll|\mathrm{\Delta}n_{\mathrm{NL}}|.
\end{equation}
In the opposite limit, the 1D photon gas is in the strong-interaction regime, the so-called Tonks--Girardeau regime \cite{Tonks1936, Girardeau1960, Olshanii1998, Thywissen1999, Petrov2000, Dunjko2001}. The study of this regime in a 1D waveguide geometry will be the subject of future works \cite{LebreuillyToBePublished, LebreuillyBisToBePublished}.

\subsection{Modulus-phase Bogoliubov theory of quantum fluctuations}
\label{SubSec:ModulusPhaseBogoliubovTheoryOfQuantumFluctuations}

When a gas of weakly interacting bosons presents a macroscopically occupied, i.e., condensate, state, it is well known that its quantum fluctuations may be accurately treated by means of the Bogoliubov theory of linearized fluctuations \cite{Dalfovo1999, Pitaevskii2003, Castin2001, Fetter2003}. In a 1D configuration, it is also known that the quantum fluctuations of the phase of the field operator which describes the fluid are strong and prevent the occurrence of a true condensation \cite{Popov1972, Popov1983}. Consequently, in this case, a more sophisticated version of the standard Bogoliubov theory of linearized fluctuations has to be used to treat in a proper way the quantum fluctuations of the weakly interacting Bose gas. In the following, we shall apply the extended Bogoliubov theory of dilute atomic Bose gases in reduced dimensions \cite{PetrovThesis, Mora2003, Petrov2004} to the 1D propagating geometry investigated in this work, assuming that the photon fluid is well in the weakly interacting regime \eqref{Eq:DiluteGasConditionExperiments}, as mentioned in the last paragraph of Sect.~\ref{SubSec:Quantization}. Note that this approach has been used in Ref.~\cite{Chiocchetta2013} to investigate the coherence properties of low-dimensional driven-dissipative photon fluids. As first indicated in Refs.~\cite{Gladilin2014, Altman2015}, the existence of driving and dissipation in the system requires including nonlinear corrections to the Bogoliubov theory of linearized fluctuations. This is however not the case in the present work, where (in the assumed absence of photon absorption) the dynamics of the quantum fluid of light is purely conservative, as in the atomic case.

Following Refs.~\cite{PetrovThesis, Mora2003, Petrov2004}, one starts by expressing the  quantum field operator \eqref{Eq:QuantumField} and its Hermitian conjugate in polar form:
\begin{subequations}
\label{Eq:MadelungPicture}
\begin{align}
\label{Eq:MadelungPicture1}
\hat{\mathrm{\Psi}}(\zeta,\tau)&=\mathrm{e}^{\mathrm{i}[\hat{\varphi}(\zeta,\tau)-\mu\tau/\hbar]}\,\sqrt{\hat{\rho}(\zeta,\tau)}, \\
\label{Eq:MadelungPicture2}
\hat{\mathrm{\Psi}}^{\dag}(\zeta,\tau)&=\sqrt{\hat{\rho}(\zeta,\tau)}\;\mathrm{e}^{-\mathrm{i}[\hat{\varphi}(\zeta,\tau)-\mu\tau/\hbar]},
\end{align}
\end{subequations}
where the square modulus of $\hat{\mathrm{\Psi}}(\zeta,\tau)$,
\begin{equation}
\label{Eq:Density}
\hat{\mathrm{\Psi}}^{\dag}(\zeta,\tau)\,\hat{\mathrm{\Psi}}(\zeta,\tau)=\hat{\rho}(\zeta,\tau)=\rho+\delta\hat{\rho}(\zeta,\tau),
\end{equation}
is assumed to weakly fluctuate around a uniform (that is, $\zeta$-independent) stationary (that is, $\tau$-independent) average value $\rho$ while the quantum phase operator $\hat{\varphi}(\zeta,\tau)$ is supposed to smoothly vary in space (i.e., to weakly depend on $\zeta$). In the classical phase shift $-\mu\,\tau/\hbar$, $\mu$ corresponds in the theory of Bose fluids to the chemical potential of the gas \cite{Dalfovo1999, Pitaevskii2003}. To be consistent with Eqs.~\eqref{Eq:1DCommutationRelations}, the real fields $\hat{\rho}(\zeta,\tau)$ and $\hat{\varphi}(\zeta,\tau)$ have to obey the following commutation relation:
\begin{equation}
\label{Eq:CommutationRelationDensityPhase}
[\hat{\rho}(\zeta,\tau),\hat{\varphi}(\zeta',\tau)]=\mathrm{i}\;\mathcal{N}\,\frac{\hbar\,\omega_{0}}{\varepsilon_{0}}\,\delta(\zeta-\zeta').
\end{equation}
As the phase $\hat{\varphi}(\zeta,\tau)$ is allowed to take arbitrarily large values, no hypothesis is made on the presence of a condensate state, as a result of which the modulus-phase reformulation \eqref{Eq:MadelungPicture}--\eqref{Eq:CommutationRelationDensityPhase} of the quantum field theory presented in Sect.~\ref{SubSec:Quantization} makes it possible to treat the absence of long-range order inherent to conservative 1D Bose systems \cite{Popov1972, Popov1983, PetrovThesis, Mora2003, Petrov2004}.

Substituting Eqs.~\eqref{Eq:MadelungPicture} into Eq.~\eqref{Eq:QuantumGPE} and separating the real and imaginary parts, one gets the well-known coupled continuity and Euler-type hydrodynamic equations for the total ``density'' $\hat{\rho}(\zeta,\tau)$ and the velocity field $\hbar\,\partial_{\zeta}\hat{\varphi}(\zeta,\tau)/m$ (see Refs.~\cite{PetrovThesis, Mora2003, Petrov2004}). Then, one linearizes these latter with respect to $\delta\hat{\rho}(\zeta,\tau)$ and $\partial_{\zeta}\hat{\varphi}(\zeta,\tau)$ around the classical, uniform, and stationary background profile $\hat{\rho}(\zeta,\tau)=\rho$ and $\partial_{\zeta}\hat{\varphi}(\zeta,\tau)=0$. The zero-order terms provide the expression of the chemical potential $\mu$,
\begin{equation}
\label{Eq:ClassicalGPEDensity}
\mu=E_{0}+g\,\rho,
\end{equation}
while the first-order ones give the following coupled evolution equations for the operators $\delta\hat{\rho}(\zeta,\tau)$ and $\hat{\varphi}(\zeta,\tau)$:
\begin{subequations}
\label{Eq:EvolutionEquationsFluctuations}
\begin{align}
\label{Eq:EvolutionEquationsFluctuations1}
\mathrm{i}\,\hbar\,\frac{\partial}{\partial\tau}\,\bigg(\frac{\delta\hat{\rho}}{\sqrt{\rho}}\bigg)&=-\frac{\hbar^{2}}{2\,m}\,\frac{\partial^{2}}{\partial\zeta^{2}}\,(2\,\mathrm{i}\,\sqrt{\rho}\,\hat{\varphi}), \\
\label{Eq:EvolutionEquationsFluctuations2}
\mathrm{i}\,\hbar\,\frac{\partial}{\partial\tau}\,(2\,\mathrm{i}\,\sqrt{\rho}\,\hat{\varphi})&=\bigg({-}\frac{\hbar^{2}}{2\,m}\,\frac{\partial^{2}}{\partial\zeta^{2}}+2\,g\,\rho\bigg)\,\bigg(\frac{\delta\hat{\rho}}{\sqrt{\rho}}\bigg).
\end{align}
\end{subequations}

The solutions $\delta\hat{\rho}(\zeta,\tau)$ and $\hat{\varphi}(\zeta,\tau)$ of Eqs.~\eqref{Eq:EvolutionEquationsFluctuations}, oscillating around the homogeneous and stationary background pattern such that $\hat{\rho}(\zeta,\tau)=\rho=\mathrm{const}$ and $\partial_{\zeta}\hat{\varphi}(\zeta,\tau)=0$ may be expressed as linear superpositions of plane-wave-type elementary excitations:
\begin{subequations}
\label{Eq:PlaneWaveExpansions}
\begin{align}
\label{Eq:PlaneWaveExpansions1}
\delta\hat{\rho}(\zeta,\tau)&=\sqrt{\rho}\int\frac{\mathrm{d}k}{2\pi}\,(u_{k}+v_{k})\,\mathrm{e}^{\mathrm{i}k\zeta}\,\hat{b}_{k}(\tau)+\mathrm{H.c.}, \\
\label{Eq:PlaneWaveExpansions2}
\hat{\varphi}(\zeta,\tau)&=\frac{1}{2\,\mathrm{i}\,\sqrt{\rho}}\int\frac{\mathrm{d}k}{2\pi}\,(u_{k}-v_{k})\,\mathrm{e}^{\mathrm{i}k\zeta}\,\hat{b}_{k}(\tau)+\mathrm{H.c.},
\end{align}
\end{subequations}
where ``H.c.'' means ``Hermitian conjugate''. In the plane-wave expansions \eqref{Eq:PlaneWaveExpansions}, the $\hat{b}_{k}^{\vphantom{\dag}}(\tau)$'s [$\hat{b}_{k}^{\dag}(\tau)$'s] are ladder operators annihilating (creating) at the propagation time $\tau$ photon excitations of energy
\begin{equation}
\label{Eq:BogoliubovSpectrum}
\hbar\,\omega_{k}=\sqrt{\hbar\,\mathrm{\Omega}_{k}\,(\hbar\,\mathrm{\Omega}_{k}+2\,g\,\rho)}
\end{equation}
in the plane-wave mode of wavenumber $k$ in the $\zeta$ direction, where the $k$-dependent angular frequency $\mathrm{\Omega}_{k}$ is defined as
\begin{equation}
\label{Eq:FreeParticleSpectrum}
\hbar\,\mathrm{\Omega}_{k}=\frac{\hbar^{2}\,k^{2}}{2\,m}.
\end{equation}
They harmonically evolve as
\begin{equation}
\label{Eq:HarmonicEvolution}
\hat{b}_{k}^{\vphantom{\mathrm{in}}}(\tau)=\mathrm{e}^{-\mathrm{i}\omega_{k}\tau}\,\hat{b}_{k}^{\mathrm{in}},
\end{equation}
where $\hat{b}_{k}^{\mathrm{in}}=\hat{b}_{k}^{\vphantom{\mathrm{in}}}(\tau=0)$, and obey the following Bose commutation rules at the same time $\tau$:
\begin{subequations}
\label{Eq:CommutationRelationsElementaryExcitations}
\begin{align}
\label{Eq:CommutationRelationsElementaryExcitations1}
[\hat{b}_{k^{\vphantom{\prime}}}^{\vphantom{\dag}}(\tau),\hat{b}_{k'}^{\dag}(\tau)]&=2\pi\,\mathcal{N}\,\frac{\hbar\,\omega_{0}}{\varepsilon_{0}}\,\delta(k-k'), \\
\label{Eq:CommutationRelationsElementaryExcitations2}
[\hat{b}_{k^{\vphantom{\prime}}}(\tau),\hat{b}_{k'}(\tau)]&=0.
\end{align}
\end{subequations}
Finally,
\begin{equation}
\label{Eq:BogoliubovAmplitudes}
u_{k},v_{k}=\frac{1}{2}\,\frac{\mathrm{\Omega}_{k}\pm\omega_{k}}{\sqrt{\mathrm{\Omega}_{k}\,\omega_{k}}}
\end{equation}
are the so-called Bogoliubov amplitudes, such that $u_{k}^{2}-v_{k}^{2}=1$.

The elementary excitations of the 1D quantum fluid of light of constant ``density'' $\rho$ are plane waves of wavenumber $\pm\,k$ in the $\zeta$ direction, i.e., of angular frequency $\delta\omega=\mp\,v_{0}\,k$ as $\zeta$ is nothing but, at a given position $z$ along the optical axis, the physical time $t$ multiplied by the group velocity $v_{0}$ [see the second of Eqs.~\eqref{Eq:GPECoordinates}]. Their $k$-dependent energy \eqref{Eq:BogoliubovSpectrum} [the $k$-dependent law \eqref{Eq:FreeParticleSpectrum}] corresponds to the well-known Bogoliubov (free-particle) dispersion relation of a uniform dilute Bose gas at rest. We then see that the hypothesis of small ``density'' fluctuations $\delta\hat{\rho}=\hat{\rho}-\rho$ and of slow variations of the phase $\hat{\varphi}$ is sufficient for having the Bogoliubov excitation spectrum, no matter there is or is not a macroscopically occupied, condensate, state in the system. Within the $z\longleftrightarrow t$ mapping adopted in this work, $\hbar\,\omega_{k}$ straightforwardly corresponds to a wavenumber along the radiation, $z$, axis.

The Bogoliubov dispersion law $\hbar\,\omega_{k}$ is a real function of $k$ when the effective photon mass $m$ and the two-photon interaction parameter $g$, that is, the group-velocity dispersion $D_{0}$ [see Eq.~\eqref{Eq:Mass}] and the Kerr-nonlinearity coefficient $n_{2}(\omega_{0})$ [see Eqs.~\eqref{Eq:ExternalPotential3DInteractionConstant}, \eqref{Eq:1DInteractionConstant}, and \eqref{Eq:InteractionConstant}], are of same sign: $m$ and $g$ $\gtrless0$, that is, $D_{0}$ and $n_{2}(\omega_{0})$ $\lessgtr0$. Considering from now on that it is the case, the waveguide's dispersion and the two-photon collision processes mediated by the Kerr nonlinearity of the underlying medium do not give rise to dynamically unstable behaviors in the 1D quantum fluid of light. At low excitation wavenumbers, i.e., when $\xi\,|k|\ll1$, where
\begin{equation}
\label{Eq:HealingLength}
\xi=\frac{\hbar}{\sqrt{m\,g\,\rho}}
\end{equation}
is the healing length of the light fluid, $\hbar\,\omega_{k}$ is phononlike: $\hbar\,\omega_{k}\simeq s\,|\hbar\,k|$, where
\begin{equation}
\label{Eq:SoundVelocity}
s=\sqrt{\frac{g\,\rho}{m}}=\frac{\hbar}{|m|\,\xi}
\end{equation}
is the speed of sound in the photon fluid. An experiment aiming at probing the linear part of $\hbar\,\omega_{k}$ and, in turn, at measuring $s$ in a 1D-waveguide geometry is presently in progress \cite{BiasiToBePublished}. In the opposite regime, i.e., when $\xi\,|k|\gg1$, $\hbar\,\omega_{k}$ approaches the particlelike dispersion relation \eqref{Eq:FreeParticleSpectrum}: $\hbar\,\omega_{k}\simeq|\hbar\,\mathrm{\Omega}_{k}|+|g|\,\rho$; in the present context, the Hartree interaction term $|g|\,\rho$ corresponds to the standard modification of the propagation constant due to the optical nonlinearity of the underlying medium \cite{Larre2015a}.

\section{Light coherence in response to quantum quenches in the Kerr nonlinearity}
\label{Sec:LightCoherenceInResponseToQuantumQuenchesInTheKerrNonlinearity}

Up to now, we endeavoured to review in Sect.~\ref{Sec:ClassicalWaveEquation} the paraxial propagation of a quasimonochromatic electromagnetic wave in a 1D nonlinear waveguide, to quantize in Sect.~\ref{SubSec:Quantization} the corresponding 1D classical field theory, and finally to describe in Sect.~\ref{SubSec:ModulusPhaseBogoliubovTheoryOfQuantumFluctuations} the evolution of the quantum fluctuations of the 1D optical field in the bulk of the waveguide, assumed to be weakly nonlinear. Making use of these results, we investigate in the present section the light beam's coherence properties resulting from the existence of the air--nonlinear-dielectric interfaces which delimit the waveguide in the $z$ direction.

Having in mind the reformulation of the propagation of the optical field in the increasing-$z$ direction in terms of a temporal evolution (cf.~Sect.~\ref{SubSubSec:GrossPitaevskiiTypeTimeEvolution}), it is straightforward to see evidence that the propagating configuration investigated in the present work constitutes a very simple realization of a pair of quantum quenches of the Hamiltonian of the full 1D optical system in the nonlinear photon-photon interaction parameter $g$: As the optical nonlinearity is nonzero only inside the Kerr material constituting the waveguide, the first (second) quench occurs at the entrance (exit) face of the nonlinear medium, where the value of the two-photon interaction constant $g$ suddenly jumps from $0$ to $\propto n_{2}(\omega_{0})\neq0$ [from $\propto n_{2}(\omega_{0})\neq0$ to $0$]; see Fig.~\ref{Fig:ExperimentalSetup}.

While the present study focuses on the dilute-gas limit \eqref{Eq:DiluteGasConditionExperiments}, regime within which the quantum fluctuations of the 1D photon fluid are accurately described by the modulus-phase Bogoliubov theory of dilute Bose gases (cf.~Sect.~\ref{SubSec:ModulusPhaseBogoliubovTheoryOfQuantumFluctuations}), application of the general 1D quantum theory presented in Sect.~\ref{SubSec:Quantization} to the strongly interacting, Tonks--Girardeau, regime is possible and will be the subject of forthcoming publications \cite{LebreuillyToBePublished, LebreuillyBisToBePublished}.

After having presented in Sect.~\ref{SubSec:PhysicalSituation} the physical situation, we shall provide in Sect.~\ref{SubSec:TimeEvolutionOfTheQuantumFluctuations} an analytical description of the evolution of the quantum fluctuations of the optical field accounting for the presence of the air-waveguide and waveguide-air interfaces. Then, in Sect.~\ref{SubSec:FirstOrderCoherenceOfTheTransmittedLight}, we shall study the coherence properties of the transmitted beam of light, in response to the sudden quenches in the photon-photon interaction constant.

\subsection{Physical situation}
\label{SubSec:PhysicalSituation}

\begin{figure}[t!]
\includegraphics[width=\linewidth]{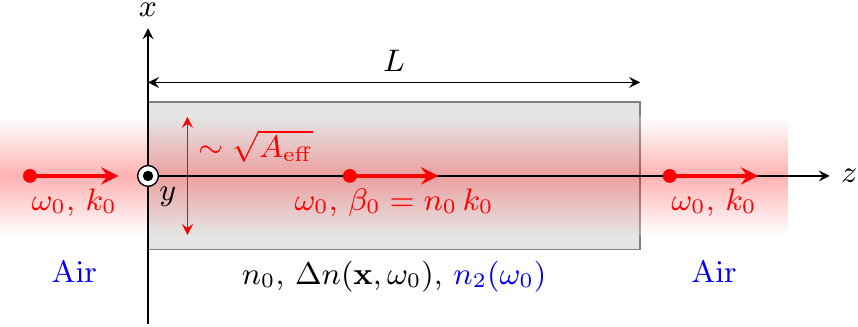}
\caption{Sketch of the considered experimental configuration. A laser beam of angular frequency $\omega_{0}$ and propagation constant $k_{0}$ in the positive-$z$ direction is sent in a 1D nonlinear optical waveguide of length $L$ along the $z$ axis, background refractive index $n_{0}$, optical confinement $\mathrm{\Delta}n(\mathbf{x},\omega_{0})$, and Kerr coefficient $n_{2}(\omega_{0})$. At the entrance (exit) face of the nonlinear material, i.e., at $z=0$ ($z=L$), the photon-photon interaction constant suddenly jumps from zero to $\propto n_{2}(\omega_{0})$ [from $\propto n_{2}(\omega_{0})$ to zero]; this results in a pair of quenches of the full system's quantum Hamiltonian. The degree of first-order coherence \eqref{Eq:DOFOC} of the transmitted light, i.e., emerging from the $z=L$ exit face of the waveguide, reveals a loss of coherence in the 1D quantum fluid of light.}
\label{Fig:ExperimentalSetup}
\end{figure}

A sketch of the above-discussed configuration is drawn in Fig.~\ref{Fig:ExperimentalSetup}. The 1D nonlinear optical waveguide is encompassed between $z=0$ (defining its entrance face) and $z=L$ (corresponding to its exit face). Its group-velocity dispersion $D_{0}=D(\omega_{0})$ and its Kerr coefficient $n_{2}(\omega_{0})$ at the laser pump's angular frequency $\omega_{0}$ are assumed to be of same sign to prevent the occurrence of dynamical instabilities (see the last paragraph of Sect.~\ref{SubSec:ModulusPhaseBogoliubovTheoryOfQuantumFluctuations}) in the 1D photon fluid, supposed to be well within the weakly interacting regime \eqref{Eq:DiluteGasConditionExperiments}. The quasimonochromatic beam which illuminates the $z=0$ entrance face of the waveguide is assumed to have a wide top-hat spatial profile in the $\mathbf{x}=(x,y)$ plane as well as a constant power all along the optical axis before entering the waveguide so that it can be legitimately seen as an infinite plane wave propagating without amplitude attenuations in the increasing-$z$ direction.

Back-propagating waves originating from reflection on the entrance (at $z=0$) and the back (at $z=L$) faces of the waveguide would spoil the reformulation of light propagation in the positive-$z$ direction in terms of a time evolution. To avoid dealing with their existence, we assume that the $z=0,L$ interfaces are treated with a perfect antireflection coating. As its characteristic thickness in the $z$ direction (of the order of a few optical wavelengths) is typically much smaller than all the other lengths involved in the investigated configuration, its effect on light transmission can be summarized as simple boundary conditions guaranteeing the conservation of the electromagnetic-energy flux, i.e., the Poynting vector, at the $z=0,L$ surfaces:
\begin{subequations}
\label{Eq:ContinuityPoyntingVector}
\begin{align}
\label{Eq:ContinuityPoyntingVector1}
\hat{\mathcal{E}}(\mathbf{x},z=0^{-},t)&=\sqrt{n_{0}}\;\hat{\mathcal{E}}(\mathbf{x},z=0^{+},t), \\
\label{Eq:ContinuityPoyntingVector2}
\sqrt{n_{0}}\;\hat{\mathcal{E}}(\mathbf{x},z=L^{-},t)\,\mathrm{e}^{\mathrm{i}\beta_{0}L}&=\hat{\mathcal{E}}(\mathbf{x},z=L^{+},t)\,\mathrm{e}^{\mathrm{i}k_{0}L},
\end{align}
\end{subequations}
where $\hat{\mathcal{E}}(\mathbf{x},z,t)$ denotes the quantized complex amplitude of the electric field in air ($z<0$ or $z>L$) or in the waveguide ($0<z<L$). Equations \eqref{Eq:ContinuityPoyntingVector} allow for matching the quantum fluctuations of the optical field in air to the ones propagating in the nonlinear waveguide (see Sect.~\ref{SubSubSec:MatchingAtTheInterfacesAndScatteringMatrix}).

\subsection{Time evolution of the quantum fluctuations}
\label{SubSec:TimeEvolutionOfTheQuantumFluctuations}

\subsubsection{Optical field outside the waveguide}
\label{SubSubSec:OpticalFieldOutsideTheWaveguide}

Outside the waveguide (when $z<0$ or $z>L$), i.e., in air, the quantized electric field
\begin{equation}
\label{Eq:ElectricFieldAir}
\hat{E}(\mathbf{x},z,t)=\tfrac{1}{2}\,\hat{\mathcal{E}}(\mathbf{x},z,t)\,\mathrm{e}^{\mathrm{i}(k_{0}z-\omega_{0}t)}+\mathrm{H.c.}
\end{equation}
of the light wave (one recalls that $k_{0}=\omega_{0}/c_{0}$ denotes its propagation constant in the positive-$z$ direction in air) may be expressed in terms of a quantized amplitude $\hat{\mathcal{E}}(\mathbf{x},z,t)$ that can be shown to admit, at first order in the paraxial and slowly-varying-envelope approximations considered in this work, the following expansion \cite{Larre2015a}:
\begin{equation}
\label{Eq:3DEnvelopeAir}
\hat{\mathcal{E}}(\mathbf{x},\zeta,\tau)=[\mathcal{E}_{0}+\delta\hat{\mathcal{E}}(\mathbf{x},\zeta,\tau)]\,\mathrm{e}^{\mathrm{i}\varphi(\tau)},
\end{equation}
written as a function of the coordinates $\zeta$ and $\tau$ defined in Eqs.~\eqref{Eq:GPECoordinates}. In Eq.~\eqref{Eq:3DEnvelopeAir}, the classical amplitude $\mathcal{E}_{0}$, supposed constant [that is, independent on $\mathbf{x}$, $z$, and $t$ (see the first paragraph of Sect.~\ref{SubSec:PhysicalSituation}), and so on $\mathbf{x}$, $\tau$, and $\zeta$, respectively] and real, corresponds to the coherent, classical, component of the envelope of the light wave's electric field in air, $\varphi(\tau)$ is a global phase that is assumed piecewise constant, null before entering the medium ($z<0$, i.e., $\tau<0$) and equal to $\varphi_{>}$ after exiting it ($z>L$, i.e., $\tau>L/v_{0}$), and the field operator $\delta\hat{\mathcal{E}}(\mathbf{x},\zeta,\tau)$ is a small quantum correction to the classical amplitude $\mathcal{E}_{0}$ that may be shown to satisfy the plane-wave decomposition
\begin{equation}
\label{Eq:3DPlaneWaveExpansionAir}
\delta\hat{\mathcal{E}}(\mathbf{x},\zeta,\tau)=\sqrt{\frac{v_{0}}{c_{0}}}\int\frac{\mathrm{d}^{2}q\,\mathrm{d}k}{(2\pi)^{3}}\,\mathrm{e}^{\mathrm{i}(\mathbf{q}\cdot\mathbf{x}+k\zeta)}\,\hat{\alpha}_{\mathbf{q},k}(\tau).
\end{equation}
In this expansion, the ratio $v_{0}/c_{0}$ under the square root originates from the fact that we use the same definition for the variables $\zeta$ and $\tau$ [cf.~Eqs.~\eqref{Eq:GPECoordinates}] irrespective of whether one is outside or inside the medium, although it would have been more natural to define $\zeta$ and $\tau$ in air as $\zeta=c_{0}\,t-z$ and $\tau=z/c_{0}$; as explained in Ref.~\cite{Larre2015a}, this will facilitate the matching of the fields at the air-dielectric interfaces (see Sect.~\ref{SubSubSec:MatchingAtTheInterfacesAndScatteringMatrix}). On the other hand, the ladder operators $\hat{\alpha}_{\mathbf{q},k}(\tau)$ are deduced from the usual photon-annihilation operators $\hat{\gamma}_{\mathbf{q},q_{z}}$ in the state of wavevector $(\mathbf{q},q_{z})$, where $\mathbf{q}=(q_{x},q_{y})$ [in passing, $\mathrm{d}^{2}q=\mathrm{d}q_{x}\,\mathrm{d}q_{y}$ in Eq.~\eqref{Eq:3DPlaneWaveExpansionAir}], as
\begin{equation}
\label{Eq:ParaxialRealPhotons}
\hat{\alpha}_{\mathbf{q},k}(\tau)=\mathrm{i}\,\sqrt{2\,\frac{v_{0}}{c_{0}}\,\frac{\hbar\,\omega_{0}}{\varepsilon_{0}}}\,\mathrm{e}^{\mathrm{i}v_{0}[\delta q_{z}(\mathbf{q},k)+k]\tau}\,\hat{\gamma}_{\mathbf{q},k_{0}+\delta q_{z}(\mathbf{q},k)},
\end{equation}
where
\begin{equation}
\label{Eq:PhotonDispersionRelationAir}
\delta q_{z}(\mathbf{q},k)=-\frac{\mathbf{q}^{2}}{2\,k_{0}}-\frac{v_{0}}{c_{0}}\,k
\end{equation}
is the photon dispersion relation $\omega(\mathbf{q},q_{z})=c_{0}\,\sqrt{\mathbf{q}^{2}+q_{z}^{2}}$ in air expanded at first order in $\mathbf{q}^{2}$ ($|\mathbf{q}|/k_{0}\ll1$)~and~the small $z$-wavevector and frequency deviations $\delta q_{z}=q_{z}-k_{0}$ ($|\delta q_{z}|/k_{0}\ll1$) and $\delta\omega=\omega(\mathbf{q},q_{z})-\omega_{0}=-v_{0}\,k$ ($|\delta\omega|/\omega_{0}=v_{0}\,|k|/\omega_{0}\ll1$), i.e., at first order in the paraxial and slowly-varying-envelope approximations; by means of Eqs.~\eqref{Eq:ParaxialRealPhotons}, \eqref{Eq:PhotonDispersionRelationAir}, and the well-known commutation rules of the $\hat{\gamma}_{\mathbf{q},q_{z}}$'s, one shows that the paraxial-photon-destruction operators $\hat{\alpha}_{\mathbf{q},k}(\tau)$ are subject to the equal-$\tau$ Bose commutation relations
\begin{subequations}
\label{Eq:3DCommutationRelationsAir}
\begin{align}
\notag
[\hat{\alpha}_{\mathbf{q}^{\vphantom{\prime}},k^{\vphantom{\prime}}}^{\vphantom{\dag}}(\tau),\hat{\alpha}_{\mathbf{q}',k'}^{\dag}(\tau)]&\left.=(2\pi)^{3}\,\mathcal{N}_{\mathrm{vac}}\,\frac{\hbar\,\omega_{0}}{\varepsilon_{0}}\right. \\
\label{Eq:3DCommutationRelationsAir1}
&\left.\hphantom{=}\times\delta^{(2)}(\mathbf{q}-\mathbf{q}')\,\delta(k-k'),\right. \\
\label{Eq:3DCommutationRelationsAir2}
[\hat{\alpha}_{\mathbf{q},k}(\tau),\hat{\alpha}_{\mathbf{q}',k'}(\tau)]&\left.=0,\right.
\end{align}
\end{subequations}
where $\mathcal{N}_{\mathrm{vac}}=2$ is the vacuum value of the normalization constant \eqref{Eq:NormalizationConstant}.

The 1D reduction of the above-sketched 3D description of the quantum fluctuations of the electric field in air is naturally accomplished by projecting the 3D quantum field operator \eqref{Eq:3DEnvelopeAir} onto the state describing the transverse motion, in the same way as was done in Sect.~\ref{SubSec:Quantization} [see Eq.~\eqref{Eq:QuantumField}]. To facilitate the matching \eqref{Eq:ContinuityPoyntingVector} of the Poynting vector at the front and the back faces of the nonlinear waveguide (see Sect.~\ref{SubSubSec:MatchingAtTheInterfacesAndScatteringMatrix}), we project Eq.~\eqref{Eq:3DEnvelopeAir} onto the fundamental transverse modal function $\mathrm{\Phi}_{0}(\mathbf{x})$ of the waveguide, exactly as in Eq.~\eqref{Eq:QuantumField}, which yields
\begin{equation}
\label{Eq:1DEnvelopeAir}
\hat{\mathrm{\Psi}}(\zeta,\tau)=[\sqrt{\rho_{\mathrm{air}}}+\delta\hat{\mathrm{\Psi}}(\zeta,\tau)]\,\mathrm{e}^{\mathrm{i}\varphi(\tau)},
\end{equation}
where $\sqrt{\rho_{\mathrm{air}}}=\int\mathrm{d}^{2}x\;\mathrm{\Phi}_{0}^{\ast}(\mathbf{x})\,\mathcal{E}_{0}^{\vphantom{\ast}}$ and
\begin{equation}
\label{Eq:1DPlaneWaveExpansionAir}
\delta\hat{\mathrm{\Psi}}(\zeta,\tau)=\sqrt{\frac{v_{0}}{c_{0}}}\int\frac{\mathrm{d}k}{2\pi}\,\mathrm{e}^{\mathrm{i}k\zeta}\,\hat{a}_{k}(\tau).
\end{equation}
In Eq.~\eqref{Eq:1DPlaneWaveExpansionAir}, the $\hat{a}_{k}(\tau)$'s are defined as
\begin{equation}
\label{Eq:1DAnnihilationOperatorsAir}
\hat{a}_{k}(\tau)=\int\frac{\mathrm{d}^{2}q}{(2\pi)^{2}}\,\tilde{\mathrm{\Phi}}_{0}^{\ast}(\mathbf{q})\,\hat{\alpha}_{\mathbf{q},k}(\tau),
\end{equation}
where $\tilde{\mathrm{\Phi}}_{0}(\mathbf{q})=\int\mathrm{d}^{2}x\;\mathrm{\Phi}_{0}(\mathbf{x})\,\mathrm{e}^{-\mathrm{i}\mathbf{q}\cdot\mathbf{x}}$ is the Fourier transform of $\mathrm{\Phi}_{0}(\mathbf{x})$; making use of Eqs.~\eqref{Eq:3DCommutationRelationsAir} and of the normalization condition $\int\mathrm{d}^{2}q\;|\tilde{\mathrm{\Phi}}_{0}(\mathbf{q})|^{2}/(2\pi)^{2}=\int\mathrm{d}^{2}x\;|\mathrm{\Phi}_{0}(\mathbf{x})|^{2}=1$, one easily shows that they have to satisfy the same-$\tau$ commutation rules
\begin{subequations}
\label{Eq:1DCommutationRelationsAir}
\begin{align}
\label{Eq:1DCommutationRelationsAir1}
[\hat{a}_{k^{\vphantom{\prime}}}^{\vphantom{\dag}}(\tau),\hat{a}_{k'}^{\dag}(\tau)]&=2\pi\,\mathcal{N}_{\mathrm{vac}}\,\frac{\hbar\,\omega_{0}}{\varepsilon_{0}}\,\delta(k-k'), \\
\label{Eq:1DCommutationRelationsAir2}
[\hat{a}_{k}(\tau),\hat{a}_{k'}(\tau)]&=0.
\end{align}
\end{subequations}

In order to be consistent with the modulus-phase formulation (\ref{Eq:MadelungPicture}, \ref{Eq:Density}) of the 1D optical field $\hat{\mathrm{\Psi}}(\zeta,\tau)$ propagating in the waveguide, we can reexpress its counterpart in air, given by Eq.~\eqref{Eq:1DEnvelopeAir}, in the following polar form:
\begin{equation}
\label{Eq:MadelungPictureAir}
\hat{\mathrm{\Psi}}(\zeta,\tau)=\mathrm{e}^{\mathrm{i}[\hat{\varphi}(\zeta,\tau)+\varphi(\tau)]}\,\sqrt{\rho_{\mathrm{air}}+\delta\hat{\rho}(\zeta,\tau)}.
\end{equation}
Requiring that the quantum field operators $\delta\hat{\rho}(\zeta,\tau)$ and $\hat{\varphi}(\zeta,\tau)$ have to be linear combinations of the $\hat{a}_{k}(\tau)$'s, one expands Eq.~\eqref{Eq:MadelungPictureAir} up to first order in powers of $\delta\hat{\rho}$ and $\hat{\varphi}$ and identifies the resulting equation with Eq.~\eqref{Eq:1DEnvelopeAir}, which is by nature linear in the $\hat{a}_{k}(\tau)$'s [see Eq.~\eqref{Eq:1DPlaneWaveExpansionAir}]; this leads to
\begin{subequations}
\label{Eq:DensityAir}
\begin{align}
\label{Eq:DensityAir1}
\delta\hat{\rho}(\zeta,\tau)&=\sqrt{\rho_{\mathrm{air}}}\;\delta\hat{\mathrm{\Psi}}(\zeta,\tau)+\mathrm{H.c.} \\
\label{Eq:DensityAir2}
&=\sqrt{\frac{v_{0}\,\rho_{\mathrm{air}}}{c_{0}}}\int\frac{\mathrm{d}k}{2\pi}\,\mathrm{e}^{\mathrm{i}k\zeta}\,\hat{a}_{k}(\tau)+\mathrm{H.c.}
\end{align}
\end{subequations}
and
\begin{subequations}
\label{Eq:PhaseAir}
\begin{align}
\label{Eq:PhaseAir1}
\hat{\varphi}(\zeta,\tau)&=\frac{1}{2\,\mathrm{i}\,\sqrt{\rho_{\mathrm{air}}}}\,\delta\hat{\mathrm{\Psi}}(\zeta,\tau)+\mathrm{H.c.} \\
\label{Eq:PhaseAir2}
&=\frac{1}{2\,\mathrm{i}}\,\sqrt{\frac{v_{0}}{c_{0}\,\rho_{\mathrm{air}}}}\int\frac{\mathrm{d}k}{2\pi}\,\mathrm{e}^{\mathrm{i}k\zeta}\,\hat{a}_{k}(\tau)+\mathrm{H.c.}
\end{align}
\end{subequations}

We end up with formulas for the quantum operators in air very similar to the ones we have derived in Sect.~\ref{SubSec:ModulusPhaseBogoliubovTheoryOfQuantumFluctuations} in the case of the 1D nonlinear optical waveguide. The differences come from the fact that, in air, the refractive index equals one and the Kerr coefficient is zero, explaining the $\mathcal{N}_{\mathrm{vac}}=2$ in the commutation relation \eqref{Eq:CommutationRelationsElementaryExcitations1}, as well as the absence, in the plane-wave expansions \eqref{Eq:DensityAir2} and \eqref{Eq:PhaseAir2}, of the $k$-dependent weights $u_{k}\pm v_{k}=(\mathrm{\Omega}_{k}/\omega_{k})^{\pm1/2}$ present in Eqs.~\eqref{Eq:PlaneWaveExpansions}, here equal to one (indeed, $\omega_{k}\equiv\mathrm{\Omega}_{k}$ when $g\equiv0$). Finally, the square-root factor $\sqrt{v_{0}/c_{0}}$ in Eqs.~\eqref{Eq:DensityAir2} and \eqref{Eq:PhaseAir2} originates from Eq.~\eqref{Eq:3DPlaneWaveExpansionAir}.

\subsubsection{Optical field inside the waveguide}
\label{SubSubSec:OpticalFieldInsideTheWaveguide}

Inside the waveguide (i.e., in the $0<z<L$ region), the spatiotemporal evolution of the optical field is ruled by the quantum Gross--Pitaevskii equation \eqref{Eq:QuantumGPE}. In the case where the average power $\mathcal{P}=\frac{1}{2}\,c_{0}\,\varepsilon_{0}\,n_{0}\,\rho$ of the light beam in the dielectric assumes a homogeneous and stationary (that is, $\zeta$- and $\tau$- independent) profile all along the 1D propagation between the entrance (at $z=0$) and the exit (at $z=L$) of the nonlinear waveguide, the photon field $\hat{\mathrm{\Psi}}(\zeta,\tau)$ satisfying Eq.~\eqref{Eq:QuantumGPE} may be written in the polar form (\ref{Eq:MadelungPicture1}, \ref{Eq:Density}), with the fluctuating operators $\delta\hat{\rho}(\zeta,\tau)=\hat{\rho}(\zeta,\tau)-\rho$ and $\hat{\varphi}(\zeta,\tau)$ given by Eqs.~\eqref{Eq:PlaneWaveExpansions} in the weak-Kerr-nonlinearity regime \eqref{Eq:DiluteGasConditionExperiments} considered in this work.

\subsubsection{Matching at the interfaces and scattering matrix}
\label{SubSubSec:MatchingAtTheInterfacesAndScatteringMatrix}

The way in which the output quantum state of the light beam, after propagation in the nonlinear medium ($z>L$, i.e., $\tau>L/v_{0}$), depends on the input one, before entering the waveguide ($z<0$, i.e., $\tau<0$), is found by matching the quantum fields \eqref{Eq:MadelungPicture1} and \eqref{Eq:MadelungPictureAir} through Eqs.~\eqref{Eq:ContinuityPoyntingVector}, i.e., after projecting those equations onto $\mathrm{\Phi}_{0}(\mathbf{x})$, through
\begin{subequations}
\label{Eq:ContinuityPoyntingVector1D}
\begin{align}
\label{Eq:ContinuityPoyntingVector1D1}
\hat{\mathrm{\Psi}}(\zeta,\tau=0^{-})&=\sqrt{n_{0}}\;\hat{\mathrm{\Psi}}(\zeta,\tau=0^{+}), \\
\label{Eq:ContinuityPoyntingVector1D2}
\sqrt{n_{0}}\;\hat{\mathrm{\Psi}}(\zeta,\tau=\mathrm{\Delta}\tau^{-})\,\mathrm{e}^{\mathrm{i}\beta_{0}L}&=\hat{\mathrm{\Psi}}(\zeta,\tau=\mathrm{\Delta}\tau^{+})\,\mathrm{e}^{\mathrm{i}k_{0}L},
\end{align}
\end{subequations}
where we introduced, for future convenience, the notation $\mathrm{\Delta}\tau=L/v_{0}$, which corresponds to the time spent in the nonlinear waveguide of length $L$ by a photon wavepacket propagating in the positive-$z$ direction at constant velocity $v_{0}$.

At the mean-field level, the continuity equations \eqref{Eq:ContinuityPoyntingVector1D} of the Poynting vector at the $z=0,L$ interfaces entails (i) that the background ``densities'' $\rho_{\mathrm{air}}$ and $\rho$ of the fluid of light outside and inside the waveguide have to be related through
\begin{equation}
\label{Eq:Densities<And>}
\rho_{\mathrm{air}}=n_{0}\,\rho
\end{equation}
and (ii) that the classical phase $\varphi(\tau>\mathrm{\Delta}\tau)=\varphi_{>}=\mathrm{const}$ of the optical field after exiting the medium must be given by
\begin{equation}
\label{Eq:Phase>}
\varphi_{>}=(\beta_{0}-k_{0})\,L-\frac{\mu\,\mathrm{\Delta}\tau}{\hbar}=\Big(\beta_{0}-k_{0}-\frac{\mu}{\hbar\,v_{0}}\Big)\,L.
\end{equation}

We now move to the matching of the quantum fluctuations, defining, first, the input and the output mode operators $\hat{a}_{k}^{\mathrm{in}}$ and $\hat{a}_{k}^{\mathrm{out}}$ of the scattering problem as
\begin{equation}
\label{Eq:InOutLadderOperators}
\hat{a}_{k}^{\mathrm{in}}=\hat{a}_{k}^{\vphantom{\mathrm{in}}}(\tau=0^{-})\quad\text{and}\quad\hat{a}_{k}^{\mathrm{out}}=\hat{a}_{k}^{\vphantom{\mathrm{out}}}(\tau=\mathrm{\Delta}\tau^{+}),
\end{equation}
respectively. At the entrance face of the waveguide, that is, at $\tau=0$, Eq.~\eqref{Eq:ContinuityPoyntingVector1D1} yields
\begin{equation}
\label{Eq:InMatching}
\begin{bmatrix}
\hat{b}_{k}^{\mathrm{in}} \\ (\hat{b}_{-k}^{\mathrm{in}})^{\dag}
\end{bmatrix}
=S_{k}^{\mathrm{in}}
\begin{bmatrix}
\hat{a}_{k}^{\mathrm{in}} \\ (\hat{a}_{-k}^{\mathrm{in}})^{\dag}
\end{bmatrix}
,
\end{equation}
where $\hat{b}_{k}^{\mathrm{in}}=\hat{b}_{k}^{\vphantom{\mathrm{in}}}(\tau=0^{+})$ [see Eq.~\eqref{Eq:HarmonicEvolution}] and
\begin{equation}
\label{Eq:InScatteringMatrix}
S_{k}^{\mathrm{in}}=\sqrt{\frac{\mathcal{N}}{\mathcal{N}_{\mathrm{vac}}}}
\begin{bmatrix}
u_{k} & -v_{k} \\ -v_{k} & u_{k}
\end{bmatrix}
.
\end{equation}
At the exit, that is, at $\tau=\mathrm{\Delta}\tau$, Eq.~\eqref{Eq:ContinuityPoyntingVector1D2} leads to
\begin{equation}
\label{Eq:OutMatching}
\begin{bmatrix}
\hat{a}_{k}^{\mathrm{out}} \\ (\hat{a}_{-k}^{\mathrm{out}})^{\dag}
\end{bmatrix}
=S_{k}^{\mathrm{out}}
\begin{bmatrix}
\hat{b}_{k}^{\mathrm{out}} \\ (\hat{b}_{-k}^{\mathrm{out}})^{\dag}
\end{bmatrix}
,
\end{equation}
where $\hat{b}_{k}^{\mathrm{out}}=\hat{b}_{k}^{\vphantom{out}}(\tau=\mathrm{\Delta}\tau^{-})=\mathrm{e}^{-\mathrm{i}\omega_{k}\mathrm{\Delta}\tau}\,\hat{b}_{k}^{\mathrm{in}}$ [cf.~Eq.~\eqref{Eq:HarmonicEvolution}] and
\begin{equation}
\label{Eq:OutScatteringMatrix}
S_{k}^{\mathrm{out}}=\sqrt{\frac{\mathcal{N}_{\mathrm{vac}}}{\mathcal{N}}}
\begin{bmatrix}
u_{k} & v_{k} \\
v_{k} & u_{k}
\end{bmatrix}
.
\end{equation}
Substituting the above-given definition of the $\hat{b}_{k}^{\mathrm{out}}$'s and Eq.~\eqref{Eq:InMatching} into Eq.~\eqref{Eq:OutMatching}, one eventually finds that
\begin{equation}
\label{Eq:InOutMatching}
\begin{bmatrix}
\hat{a}_{k}^{\mathrm{out}} \\ (\hat{a}_{-k}^{\mathrm{out}})^{\dag}
\end{bmatrix}
=S_{k}
\begin{bmatrix}
\hat{a}_{k}^{\mathrm{in}} \\ (\hat{a}_{-k}^{\mathrm{in}})^{\dag}
\end{bmatrix}
,
\end{equation}
where
\begin{equation}
\label{Eq:ScatteringMatrix}
S_{k}^{\vphantom{\mathrm{out}}}=S_{k}^{\mathrm{out}}
\begin{bmatrix}
\mathrm{e}^{-\mathrm{i}\omega_{k}\mathrm{\Delta}\tau} & 0 \\
0 & \mathrm{e}^{\mathrm{i}\omega_{k}\mathrm{\Delta}\tau}
\end{bmatrix}
S_{k}^{\mathrm{in}}=
\begin{bmatrix}
\tilde{u}_{k}^{\vphantom{\ast}} & \tilde{v}_{k}^{\ast} \\
\tilde{v}_{k}^{\vphantom{\ast}} & \tilde{u}_{k}^{\ast}
\end{bmatrix}
\end{equation}
is the scattering matrix of the quenched propagating system. It connects the input quantum modes in air to the output ones and is defined in terms of the modified Bogoliubov amplitudes
\begin{subequations}
\label{Eq:ModifiedBogoliubovAmplitudes}
\begin{align}
\label{Eq:ModifiedBogoliubovAmplitudes1}
\tilde{u}_{k}^{\vphantom{2}}&=u_{k}^{2}\,\mathrm{e}^{-\mathrm{i}\omega_{k}\mathrm{\Delta}\tau}-v_{k}^{2}\,\mathrm{e}^{\mathrm{i}\omega_{k}\mathrm{\Delta}\tau}, \\
\label{Eq:ModifiedBogoliubovAmplitudes2}
\tilde{v}_{k}&=u_{k}\,v_{k}\,\big(\mathrm{e}^{-\mathrm{i}\omega_{k}\mathrm{\Delta}\tau}-\mathrm{e}^{\mathrm{i}\omega_{k}\mathrm{\Delta}\tau}\big).
\end{align}
\end{subequations}
In the $\mathrm{\Delta}\tau\to0$ limit (that is, in the absence of waveguide), $\tilde{u}_{k}\to1$ (since $u_{k}^{2}-v_{k}^{2}=1$), $\tilde{v}_{k}\to0$, and so the scattering matrix $S_{k}$ tends to the $2\times2$ identity matrix, leading to $\hat{a}_{k}^{\mathrm{out}}\to\hat{a}_{k}^{\mathrm{in}}$, as it has to be in such a limit.

For a perfectly coherent and monochromatic incident beam, which one assumes here, the quantum modes of the incident field are in the vacuum state $|\mathrm{vac}\rangle$ of the input operators $\hat{a}_{k}^{\mathrm{in}}$, defined as $\hat{a}_{k}^{\mathrm{in}}\,|\mathrm{vac}\rangle=0$ for all $k$, that is, for all optical angular frequency $\omega$ different from the laser pump's one $\omega_{0}$ [see the discussion between Eqs.~\eqref{Eq:PhotonDispersionRelationAir} and \eqref{Eq:3DCommutationRelationsAir}]. As a consequence, denoting the quantum average in the vacuum state $|\mathrm{vac}\rangle$ as $\langle\cdot\rangle=\langle\mathrm{vac}|\cdot|\mathrm{vac}\rangle$, one has
\begin{equation}
\label{Eq:AveragesOfInterest1}
\langle\hat{a}_{k^{\vphantom{\prime}}}^{\mathrm{in}}\,\hat{a}_{k'}^{\mathrm{in}}\rangle=\langle(\hat{a}_{k^{\vphantom{\prime}}}^{\mathrm{in}})^{\dag}\,\hat{a}_{k'}^{\mathrm{in}}\rangle=0
\end{equation}
and, by virtue of the commutation relation \eqref{Eq:1DCommutationRelationsAir1},
\begin{equation}
\label{Eq:AveragesOfInterest2}
\langle\hat{a}_{k^{\vphantom{\prime}}}^{\mathrm{in}}\,(\hat{a}_{k'}^{\mathrm{in}})^{\dag}\rangle=2\pi\,\mathcal{N}_{\mathrm{vac}}\,\frac{\hbar\,\omega_{0}}{\varepsilon_{0}}\,\delta(k-k').
\end{equation}
The statistical properties of the light beam emerging from the waveguide are fully obtained by considering these identities as the initial conditions of the propagating problem. For example, the momentum-space correlation functions
\begin{equation}
\label{Eq:MomentumDistributionDefinition}
n_{k,k'}=\langle\hat{a}_{k^{\vphantom{\prime}}}^{\dag}(\mathrm{\Delta}\tau^{+})\,\hat{a}_{k'}^{\vphantom{\dag}}(\mathrm{\Delta}\tau^{+})\rangle=\langle(\hat{a}_{k^{\vphantom{\prime}}}^{\mathrm{out}})^{\dag}\,\hat{a}_{k'}^{\mathrm{out}}\rangle
\end{equation}
and
\begin{equation}
\label{Eq:AnomalousAverageDefinition}
m_{k,k'}=\langle\hat{a}_{k}(\mathrm{\Delta}\tau^{+})\,\hat{a}_{-k'}(\mathrm{\Delta}\tau^{+})\rangle=\langle\hat{a}_{k^{\vphantom{\prime}}}^{\mathrm{out}}\,\hat{a}_{-k'}^{\mathrm{out}}\rangle
\end{equation}
at the back face of the nonlinear waveguide\footnote{When $k'=k$, the correlation functions \eqref{Eq:MomentumDistributionDefinition} and \eqref{Eq:AnomalousAverageDefinition} correspond to the so-called normal and anomalous averages in the physics of atomic Bose gases. Note also that $\int\mathrm{d}k'\,n_{k,k'}/(2\pi)$ is directly proportional to the excitation momentum distribution at the exit of the waveguide.} are obtained by combining Eqs.~\eqref{Eq:InOutMatching}, \eqref{Eq:AveragesOfInterest1}, and \eqref{Eq:AveragesOfInterest2}, which yields
\begin{equation}
\label{Eq:MomentumDistributionResult}
n_{k,k'}=2\pi\,\mathcal{N}_{\mathrm{vac}}\,\frac{\hbar\,\omega_{0}}{\varepsilon_{0}}\,|\tilde{v}_{k}|^{2}\,\delta(k-k')
\end{equation}
and
\begin{equation}
\label{Eq:AnomalousAverageResult}
m_{k,k'}=2\pi\,\mathcal{N}_{\mathrm{vac}}\,\frac{\hbar\,\omega_{0}}{\varepsilon_{0}}\,\tilde{u}_{k}^{\vphantom{\ast}}\,\tilde{v}_{k}^{\ast}\,\delta(k-k').
\end{equation}
These results, and most particularly Eq.~\eqref{Eq:MomentumDistributionResult}, will be useful in the next section.

\subsection{First-order coherence of the transmitted light}
\label{SubSec:FirstOrderCoherenceOfTheTransmittedLight}

In this section, we propose to investigate the first-order coherence properties of the light emerging from the waveguide, in response to the pair of quenches in the Kerr nonlinearity at $z=0$ (i.e., $\tau=0$) and $z=L$ (i.e., $\tau=\mathrm{\Delta}\tau$).

From a theoretical (experimental) point of view, this amounts to calculate (measure) the degree of first-order coherence \cite{Loudon1973}
\begin{equation}
\label{Eq:DOFOC}
g_{1}(\zeta,\zeta')=\frac{G_{1}(\zeta,\zeta')}{\sqrt{G_{1}(\zeta,\zeta)\,G_{1}(\zeta',\zeta')}}
\end{equation}
after propagation of the 1D optical field $\hat{\mathrm{\Psi}}(\zeta,\tau)$ through the medium, say, at the instant $\tau=\mathrm{\Delta}\tau^{+}$---that is, at the $z=L^{+}$ exit face of the waveguide\footnote{The interesting quantum-quench physics featured by the $g_{1}$ function, originating from the presence of the air-nonlinear medium interfaces, may be entirely captured at the very back face of the waveguide. Afterwards, in the $z>L$ region of space, the (scalar) electric field of the light wave freely propagates in air. Its quantum state at an observation point $(\mathbf{x},z)$ close to the direction of propagation (that is, in the paraxial approximation) and at a distance $z-L\gg1/k_{0}$ reasonably far from the $z=L^{+}$ planar aperture of the optical waveguide may be deduced from the knowledge of the field radiated by the aperture by means of the Kirchhoff diffraction formula for nonmonochromatic waves (see, e.g., Ref.~\cite{Goodman2005}):
\begin{align}
\notag
&\left.\hat{\mathcal{A}}(\mathbf{x},z,t)\simeq\frac{1}{2\pi\,c_{0}\,(z-L)}\right. \\
\label{KirchhoffDiffractionFormula}
&\left.\times\int\limits_{z'=L^{+}}\mathrm{d}^{2}x'\;\frac{\partial\hat{\mathcal{A}}}{\partial t}\bigg(\mathbf{x}',z',t-\frac{\sqrt{|\mathbf{x}-\mathbf{x}'|^{2}+|z-z'|^{2}}}{c_{0}}\bigg),\right.
\end{align}
where $\hat{\mathcal{A}}(\mathbf{x},z,t)=\hat{\mathcal{E}}(\mathbf{x},z,t)\,\mathrm{e}^{\mathrm{i}k_{0}z}$.}---, as a function of the spatial coordinates $\zeta=v_{0}\,t-L^{+}$ and $\zeta'=v_{0}\,t'-L^{+}$---that is, as a function of the time parameters $t$ and $t'$. In Eq.~\eqref{Eq:DOFOC}, the $G_{1}$ function is defined as
\begin{equation}
\label{Eq:OBDM}
G_{1}(\zeta,\zeta')=\langle\hat{\mathrm{\Psi}}^{\dag}(\zeta,\mathrm{\Delta}\tau^{+})\,\hat{\mathrm{\Psi}}(\zeta',\mathrm{\Delta}\tau^{+})\rangle,
\end{equation}
where the 1D quantum field operator $\hat{\mathrm{\Psi}}(\zeta,\mathrm{\Delta}\tau^{+})$ is given by Eq.~\eqref{Eq:MadelungPictureAir} evaluated at $\tau=\mathrm{\Delta}\tau^{+}$ and the quantum average $\langle\cdot\rangle=\langle\mathrm{vac}|\cdot|\mathrm{vac}\rangle$ is taken over the $|\mathrm{vac}\rangle$ state of the input paraxial-photon-mode operators $\hat{a}_{k}^{\mathrm{in}}$, related to the output ones, the $\hat{a}_{k}^{\mathrm{out}}$'s, through Eq.~\eqref{Eq:InOutMatching}.

When $|g_{1}|=1$ ($|g_{1}|=0$, $|g_{1}|\notin\{1,0\}$), the light is said coherent (incoherent, partially coherent, respectively) \cite{Loudon1973}. In the physics of 1D atomic Bose gases, within which the 1D-confined particles may be described by a quantum field $\hat{\mathrm{\Psi}}(\zeta,\tau)$ [we use the same notations as the ones of Eq.~\eqref{Eq:OBDM}], $G_{1}(\zeta,\zeta')$ is the $(\zeta,\zeta')$ component of the one-body density matrix $\rho_{1}$ of the Bose gas: $G_{1}(\zeta,\zeta')=\langle\zeta|\rho_{1}|\zeta'\rangle$ \cite{Pitaevskii2003, Naraschewski1999}.

\subsubsection{Analytical derivation}
\label{SubSubSec:AnalyticalDerivation}

Within the modulus-phase representation \eqref{Eq:MadelungPictureAir} of the quantum optical field $\hat{\mathrm{\Psi}}(\zeta,\mathrm{\Delta}\tau^{+})$ describing the dilute 1D quantum fluid of light exiting the waveguide, the normalized correlation function \eqref{Eq:DOFOC} may be deduced from the generic formula \cite{Mora2003, Larre2013}
\begin{align}
\notag
g_{1}(\zeta,\zeta')=&\left.\exp\bigg\{{-}\frac{1}{8}\,\frac{\langle:\![\delta\hat{\rho}(\zeta,\mathrm{\Delta}\tau^{+})-\delta\hat{\rho}(\zeta',\mathrm{\Delta}\tau^{+})]^{2}\!:\rangle}{\rho_{\mathrm{air}}^{2}}\right. \\
\label{Eq:DOFOCDensityPhase}
&\left.-\frac{1}{2}\,\langle:\![\hat{\varphi}(\zeta,\mathrm{\Delta}\tau^{+})-\hat{\varphi}(\zeta',\mathrm{\Delta}\tau^{+})]^{2}\!:\rangle\bigg\},\right.
\end{align}
where the ``density'' fluctuation $\delta\hat{\rho}(\zeta,\tau)$ around $\rho_{\mathrm{air}}$ and the phase operator $\hat{\varphi}(\zeta,\tau)$ are respectively given by Eqs.~\eqref{Eq:DensityAir2} and \eqref{Eq:PhaseAir2}, and $:\cdot:$ denotes the normal-ordering operation with respect to the paraxial-photon oscillators $\hat{a}_{k}(\tau)$ in air.

At the back face of the waveguide, that is, at $\tau=\mathrm{\Delta}\tau^{+}$, the degree of first-order coherence admits a very simple formulation. On the one hand, using Eq.~\eqref{Eq:DensityAir2}, one gets
\begin{align}
\notag
&\left.-\frac{1}{8}\,\frac{\langle:\![\delta\hat{\rho}(\zeta,\mathrm{\Delta}\tau^{+})-\delta\hat{\rho}(\zeta',\mathrm{\Delta}\tau^{+})]^{2}\!:\rangle}{\rho_{\mathrm{air}}^{2}}\right. \\
\notag
&\left.\hphantom{\mathrm{SPACE}}=-\frac{v_{0}}{4\,c_{0}\,\rho_{\mathrm{air}}}\int\frac{\mathrm{d}k\,\mathrm{d}k'}{(2\pi)^{2}}\,\mathrm{Re}(n_{k,k'}+m_{k,k'})\right. \\
\label{Eq:DensityTermDOFOC}
&\left.\hphantom{\mathrm{SPACE}=}\times(\mathrm{e}^{\mathrm{i}k\zeta}-\mathrm{e}^{\mathrm{i}k\zeta'})\,(\mathrm{e}^{-\mathrm{i}k'\zeta}-\mathrm{e}^{-\mathrm{i}k'\zeta'}),\right.
\end{align}
and on the other hand, using Eq.~\eqref{Eq:PhaseAir2},
\begin{align}
\notag
&\left.-\frac{1}{2}\,\langle:\![\hat{\varphi}(\zeta,\mathrm{\Delta}\tau^{+})-\hat{\varphi}(\zeta',\mathrm{\Delta}\tau^{+})]^{2}\!:\rangle\right. \\
\notag
&\left.\hphantom{\mathrm{SPACE}}=-\frac{v_{0}}{4\,c_{0}\,\rho_{\mathrm{air}}}\int\frac{\mathrm{d}k\,\mathrm{d}k'}{(2\pi)^{2}}\,\mathrm{Re}(n_{k,k'}-m_{k,k'})\right. \\
\label{Eq:PhaseTermDOFOC}
&\left.\hphantom{\mathrm{SPACE}=}\times(\mathrm{e}^{\mathrm{i}k\zeta}-\mathrm{e}^{\mathrm{i}k\zeta'})\,(\mathrm{e}^{-\mathrm{i}k'\zeta}-\mathrm{e}^{-\mathrm{i}k'\zeta'}),\right.
\end{align}
where $n_{k,k'}$ and $m_{k,k'}$ are defined in Eqs.~\eqref{Eq:MomentumDistributionDefinition} and \eqref{Eq:AnomalousAverageDefinition}, respectively. The sum of Eqs.~\eqref{Eq:DensityTermDOFOC} and \eqref{Eq:PhaseTermDOFOC} corresponds to the argument of the exponential \eqref{Eq:DOFOCDensityPhase} defining the $g_{1}$ function at $\tau=\mathrm{\Delta}\tau^{+}$; the $m_{k,k'}$ contributions cancel out, which leaves, making use of Eq.~\eqref{Eq:MomentumDistributionResult},
\begin{align}
\notag
g_{1}(|\zeta-\zeta'|)=&\left.\exp\bigg\{{-}2\,\frac{v_{0}}{c_{0}}\,\frac{\hbar\,\omega_{0}}{\varepsilon_{0}\,\rho_{\mathrm{air}}}\right. \\
\label{Eq:DOFOCResult}
&\left.\times\int\frac{\mathrm{d}k}{2\pi}\,|\tilde{v}_{k}|^{2}\,[1-\cos(k\,|\zeta-\zeta'|)]\bigg\},\right.
\end{align}
where one redenotes $g_{1}(\zeta,\zeta')$ as $g_{1}(|\zeta-\zeta'|)$ to highlight that it is only function of the relative distance between $\zeta=v_{0}\,t-L^{+}$ and $\zeta'=v_{0}\,t'-L^{+}$ ($z=z'=L^{+}$), that is, of the time delay $|t-t'|$. This dependence is natural since we have considered that the photon fluid outside the waveguide (as well as inside, in passing) is of homogeneous, that is, $\zeta$-independent, background ``density'' $\rho_{\mathrm{air}}$.

Taking advantage of Eqs.~\eqref{Eq:BogoliubovSpectrum}--\eqref{Eq:BogoliubovAmplitudes} and \eqref{Eq:ModifiedBogoliubovAmplitudes2}, one may finally reformulate Eq.~\eqref{Eq:DOFOCResult} in the form
\begin{subequations}
\label{Eq:DOFOCResultBis}
\begin{align}
\notag
&\left.g_{1}(|\zeta-\zeta'|)\right. \\
\notag
&\left.=\exp\bigg\{{-}2\,\frac{v_{0}}{c_{0}}\,\frac{\hbar\,\omega_{0}}{\varepsilon_{0}\,\rho_{\mathrm{air}}}\,\Big(\frac{g\,\rho}{\hbar}\Big)^{2}\right. \\
\label{Eq:DOFOCResultBis1}
&\left.\hphantom{=}\times\int\frac{\mathrm{d}k}{2\pi}\,\bigg[\frac{\sin(\omega_{k}\,\mathrm{\Delta}\tau)}{\omega_{k}}\bigg]^{2}\,[1-\cos(k\,|\zeta-\zeta'|)]\bigg\}\right. \\
\notag
&\left.=\exp\bigg\{{-}\frac{8}{\pi}\,\frac{v_{0}}{c_{0}}\,\frac{\hbar\,\omega_{0}}{\varepsilon_{0}\,\rho_{\mathrm{air}}\,\xi}\right. \\
\label{Eq:DOFOCResultBis2}
&\left.\hphantom{=}\times\int\limits_{0}^{+\infty}\mathrm{d}\kappa\;\frac{\sin^{2}(\frac{1}{4}\,\kappa\,\sqrt{\kappa^{2}+4}\;r_{\mathrm{c}})}{\kappa^{2}\,(\kappa^{2}+4)}\,[1-\cos(\kappa\,r)]\bigg\},\right.
\end{align}
\end{subequations}
where
\begin{equation}
\label{Eq:NormalizedRelativeDistance}
r=\frac{|\zeta-\zeta'|}{\xi}=\frac{v_{0}\,|t-t'|}{\xi}
\end{equation}
is the relative distance between $\zeta$ and $\zeta'$ at $z=L$ normalized by the healing length $\xi=\hbar/\sqrt{m\,g\,\rho}$ ($\rho=\rho_{\mathrm{air}}/n_{0}$) of the 1D fluid of light in the nonlinear waveguide, and
\begin{equation}
\label{Eq:CriticalNormalizedRelativeDistance}
r_{\mathrm{c}}=\frac{|\zeta-\zeta'|_{\mathrm{c}}}{\xi}=\frac{v_{0}\,|t-t'|_{\mathrm{c}}}{\xi}=\frac{2\,s\,\mathrm{\Delta}\tau}{\xi},
\end{equation}
recalling that $s=\sqrt{g\,\rho/m}=\hbar/(|m|\,\xi)$ is the Bogoliubov speed of sound in the photon gas. To fully understand the rich physics hidden in Eqs.~\eqref{Eq:DOFOCResultBis}, it is useful to distinguish a few regimes.

\paragraph{1/ $r\lesssim r_{\mathrm{c}}$ regime.---}

In the $r\lesssim r_{\mathrm{c}}$ regime, i.e., according to Eqs.~\eqref{Eq:NormalizedRelativeDistance} and \eqref{Eq:CriticalNormalizedRelativeDistance}, in the case where $|\zeta-\zeta'|\lesssim|\zeta-\zeta'|_{\mathrm{c}}=2\,s\,\mathrm{\Delta}\tau$, or else when $|t-t'|\lesssim|t-t'|_{\mathrm{c}}=2\,(s/v_{0})\,\mathrm{\Delta}\tau$, the first-order correlation function \eqref{Eq:DOFOCResultBis2} may be approximated by the $r_{\mathrm{c}}$-independent---i.e., independent on $\mathrm{\Delta}\tau$, or equivalently, on the waveguide's length $L$---law
\begin{subequations}
\label{Eq:Belowrc}
\begin{align}
\notag
&\left.g_{1}^{<}(|\zeta-\zeta'|)\right. \\
\label{Eq:Belowrc1}
&\left.\hphantom{\mathrm{stal'}}=\exp\bigg[{-}\frac{4}{\pi}\,\frac{v_{0}}{c_{0}}\,\frac{\hbar\,\omega_{0}}{\varepsilon_{0}\,\rho_{\mathrm{air}}\,\xi}\int\limits_{0}^{+\infty}\mathrm{d}\kappa\;\frac{1-\cos(\kappa\,r)}{\kappa^{2}\,(\kappa^{2}+4)}\bigg]\right. \\
\label{Eq:Belowrc2}
&\left.\hphantom{\mathrm{stal'}}=\exp\bigg[{-}\frac{1}{4}\,\frac{v_{0}}{c_{0}}\,\frac{\hbar\,\omega_{0}}{\varepsilon_{0}\,\rho_{\mathrm{air}}\,\xi}\,(-1+2\,r+\mathrm{e}^{-2r})\bigg].\right.
\end{align}
\end{subequations}
Depending on the actual value of $r$, two subregimes can be identified.

\paragraph{1.a/ $r\gg1$ limit.---}

In the $r\gg1$ limit, i.e., when $|\zeta-\zeta'|\gg\xi$ ($|t-t'|\gg\xi/v_{0}$), the $r\lesssim r_{\mathrm{c}}$ approximation \eqref{Eq:Belowrc} of the $g_{1}$ function presents a thermal-like exponential behavior:
\begin{equation}
\label{Eq:InteractingThermalBehavior}
g_{1}^{<}(|\zeta-\zeta'|)\simeq\exp\bigg[{-}\pi\,\frac{|\zeta-\zeta'|}{d_{\mathrm{1D}}\,\mathrm{\Lambda}^{2}(T_{\mathrm{eff}})}\bigg],
\end{equation}
expressed as a function of the 1D number density $d_{\mathrm{1D}}$ of photons in the waveguide, which is in passing determined as the ratio of the photon flux $\mathcal{F}=\mathcal{P}/(\hbar\,\omega_{0})$, where $\mathcal{P}=\frac{1}{2}\,c_{0}\,\varepsilon_{0}\,n_{0}\,\rho=\frac{1}{2}\,c_{0}\,\varepsilon_{0}\,\rho_{\mathrm{air}}$ is the power of the beam of light, by the group velocity $v_{0}$,
\begin{equation}
\label{Eq:NumberDensity}
d_{\mathrm{1D}}=\frac{\mathcal{F}}{v_{0}}=\frac{1}{2}\,\frac{c_{0}}{v_{0}}\,\frac{\varepsilon_{0}\,n_{0}\,\rho}{\hbar\,\omega_{0}}=\frac{1}{2}\,\frac{c_{0}}{v_{0}}\,\frac{\varepsilon_{0}\,\rho_{\mathrm{air}}}{\hbar\,\omega_{0}},
\end{equation}
and in terms of the effective thermal de Broglie wavelength
\begin{equation}
\label{Eq:DeBroglieWavelength}
\mathrm{\Lambda}(T_{\mathrm{eff}})=\sqrt{\frac{2\pi\,\hbar^{2}}{|m|\,k_{\mathrm{B}}T_{\mathrm{eff}}}}
\end{equation}
evaluated at the effective quench-induced temperature
\begin{equation}
\label{Eq:EffectiveTemperature}
T_{\mathrm{eff}}=\frac{1}{k_{\mathrm{B}}}\,\frac{|g|\,\rho}{2}=\frac{1}{k_{\mathrm{B}}}\,\frac{|g|\,\rho_{\mathrm{air}}/n_{0}}{2},
\end{equation}
$k_{\mathrm{B}}$ denoting the Boltzmann constant. Notice that $\mathrm{\Lambda}(T_{\mathrm{eff}})$ given by Eq.~\eqref{Eq:DeBroglieWavelength} is nothing but a redefinition of the healing length $\xi$ of the quantum fluid of light: $\mathrm{\Lambda}(T_{\mathrm{eff}})=2\,\sqrt{\pi}\times\xi$.

\paragraph{1.b/ $r\ll1$ limit.---}

In the opposite $r\ll1$ limit, that is, when $|\zeta-\zeta'|\ll\xi$ ($|t-t'|\ll\xi/v_{0}$), the $r\lesssim r_{\mathrm{c}}$ estimation \eqref{Eq:Belowrc} of the degree of first-order coherence $g_{1}(|\zeta-\zeta'|)$ presents a Gaussian shape given by
\begin{equation}
\label{Eq:NoninteractingThermalBehavior}
g_{1}^{<}(|\zeta-\zeta'|)\simeq\exp\bigg[{-}\pi\,\frac{|\zeta-\zeta'|^{2}}{d_{\mathrm{1D}}\,\xi\,\mathrm{\Lambda}^{2}(T_{\mathrm{eff}})}\bigg],
\end{equation}
where $d_{\mathrm{1D}}$ and $\mathrm{\Lambda}(T_{\mathrm{eff}})$ are respectively defined in Eqs.~\eqref{Eq:NumberDensity} and \eqref{Eq:DeBroglieWavelength}.

\paragraph{2/ $r>r_{\mathrm{c}}$ regime.---}

To complete the analytical dissection of $g_{1}(|\zeta-\zeta'|)$, let us finally analyse the $r>r_{\mathrm{c}}$ behavior of Eq.~\eqref{Eq:DOFOCResultBis2}, i.e., the regime for which $|\zeta-\zeta'|>|\zeta-\zeta'|_{\mathrm{c}}=2\,s\,\mathrm{\Delta}\tau$, or, in the original coordinates $z$ and $t$, for which $|t-t'|>|t-t'|_{\mathrm{c}}=2\,(s/v_{0})\,\mathrm{\Delta}\tau$. In this case, the coherence function $g_{1}$ presents (small) damped oscillations around a plateau corresponding to its finite large-$r$ limit,
\begin{subequations}
\label{Eq:Aboverc}
\begin{align}
\label{Eq:Aboverc1}
g_{1}^{>}&\left.=\underset{r\to+\infty}{\lim}\,g_{\vphantom{\infty}}^{(1)}(|\zeta-\zeta'|)\right. \\
\notag
&\left.=\exp\bigg\{{-}\frac{8}{\pi}\,\frac{v_{0}}{c_{0}}\,\frac{\hbar\,\omega_{0}}{\varepsilon_{0}\,\rho_{\mathrm{air}}\,\xi}\right. \\
\label{Eq:Aboverc2}
&\left.\hphantom{=}\times\int\limits_{0}^{+\infty}\mathrm{d}\kappa\;\frac{\sin^{2}(\frac{1}{4}\,\kappa\,\sqrt{\kappa^{2}+4}\;r_{\mathrm{c}})}{\kappa^{2}\,(\kappa^{2}+4)}\bigg\}.\right.
\end{align}
\end{subequations}
Also in this case, two subregimes can be identified as a function of the value of $r_{\mathrm{c}}$.

\paragraph{2.a/ $r_{\mathrm{c}}\gg1$ limit.---}

In the $r_{\mathrm{c}}\gg1$ limiting case, corresponding to the case where the waveguide's length $L$ is such that $L\gg v_{0}\,\xi/s$, we have the following asymptotic expansion:
\begin{equation}
\label{Eq:LargeLAboverc}
\ln g_{1}^{>}=-\frac{1}{2}\,\frac{v_{0}}{c_{0}}\,\frac{\hbar\,\omega_{0}}{\varepsilon_{0}\,\rho_{\mathrm{air}}\,\xi}\,\bigg(r_{\mathrm{c}}-\frac{1}{2}\bigg)+\underset{r_{\mathrm{c}}\to+\infty}{\mathrm{o}}(1).
\end{equation}

\paragraph{2.b/ $r_{\mathrm{c}}\ll1$ limit.---}

In the opposite $r_{\mathrm{c}}\ll1$ (i.e., $L\ll v_{0}\,\xi/s$) limit, one finds
\begin{equation}
\label{Eq:LowLAboverc}
\ln g_{1}^{>}=-\frac{2}{3\,\sqrt{\pi}}\,\frac{v_{0}}{c_{0}}\,\frac{\hbar\,\omega_{0}}{\varepsilon_{0}\,\rho_{\mathrm{air}}\,\xi}\,r_{\mathrm{c}}^{3/2}+\underset{r_{\mathrm{c}}\to0}{\mathrm{o}}(1).
\end{equation}

\subsubsection{Physical discussion}
\label{SubSubSec:PhysicalDiscussion}

We are now going to discuss the analytical results established in Sect.~\ref{SubSubSec:AnalyticalDerivation}. To explain the features displayed by the degree of first-order coherence of the light beam exiting the waveguide, we will provide graphical representations of the $g_{1}$ function as well as numerical estimates of the relevant parameters of the problem making use of the optical constants listed in Tab.~\ref{Tab:ExperimentalParameters}, both for the laser beam and the 1D nonlinear optical waveguide. These parameters are inspired from silicon photonics. To put in better evidence the conservative-dynamics features, we will make the assumption of vanishing propagation losses. A quantitative condition for the accuracy of this approximation will be briefly discussed in the next-to-the-last paragraph of the present section for the most important case of one-photon absorption.

\begin{table}[t!]
\caption{Numerical parameters used for plotting Figs.~\ref{Fig:DOFOC} and \ref{Fig:MerminWagner}. $\lambda_{0}=2\pi/k_{0}=2\pi\,c_{0}/\omega_{0}$ and $\mathcal{P}=\frac{1}{2}\,c_{0}\,\varepsilon_{0}\,\rho_{\mathrm{air}}=\frac{1}{2}\,c_{0}\,\varepsilon_{0}\,n_{0}\,\rho$ are the operating wavelength in air and the average laser power. The refractive index $n_{0}$, the group velocity $v_{0}$, and the group-velocity-dispersion parameter $D_{0}$ at $\omega=\omega_{0}$ are deduced from the dispersion law $n=n(\omega)$ of silicon established in Ref.~\cite{Salzberg1957}, the Kerr coefficient $\tilde{n}_{2}(\omega_{0})=2\,n_{2}(\omega_{0})/(c_{0}\,\varepsilon_{0}\,n_{0})$, given in intensity units, may be found in Ref.~\cite{Dinu2003}, and the estimate of the effective transverse-mode area $A_{\mathrm{eff}}$ is chosen among typical values given in Ref.~\cite{Agrawal1995}. Finally, as in the text, $L$ denotes the length of the 1D nonlinear optical waveguide.}
\label{Tab:ExperimentalParameters}
\begin{center}
\begin{tabular}{ll}
\hline
\hline
\noalign{\smallskip}
Beam of light                    & 1D nonlinear optical waveguide \\
\noalign{\smallskip}\hline\noalign{\smallskip}
$\lambda_{0}=1.55~\mu\mathrm{m}$ & $n_{0}\simeq3.5$ \\
$\mathcal{P}=500~\mathrm{W}$     & $v_{0}\simeq8.3\times10^{7}~\mathrm{m}/\mathrm{s}$ \\
                                 & $D_{0}\simeq1.1\times10^{-24}~\mathrm{s}^{2}/\mathrm{m}$ \\
                                 & $\tilde{n}_{2}(\omega_{0})\simeq4.5\times10^{-18}~\mathrm{m}^{2}/\mathrm{W}$ \\
                                 & $A_{\mathrm{eff}}=1~\mu\mathrm{m}^{2}$ \\
                                 & $L=
                                   \begin{cases}
                                   1,1.5,2~\mathrm{mm} & \text{[see Fig.~\ref{Fig:DOFOC}(a)]} \vspace{-0.75mm} \\
                                   10~\mathrm{km}      & \text{[see Fig.~\ref{Fig:DOFOC}(b)]}
                                   \end{cases}$ \\
\noalign{\smallskip}
\hline
\hline
\end{tabular}
\end{center}
\end{table}

\paragraph{General behavior of the coherence function of the transmitted light.---}

\begin{figure}[t!]
\includegraphics[width=\linewidth]{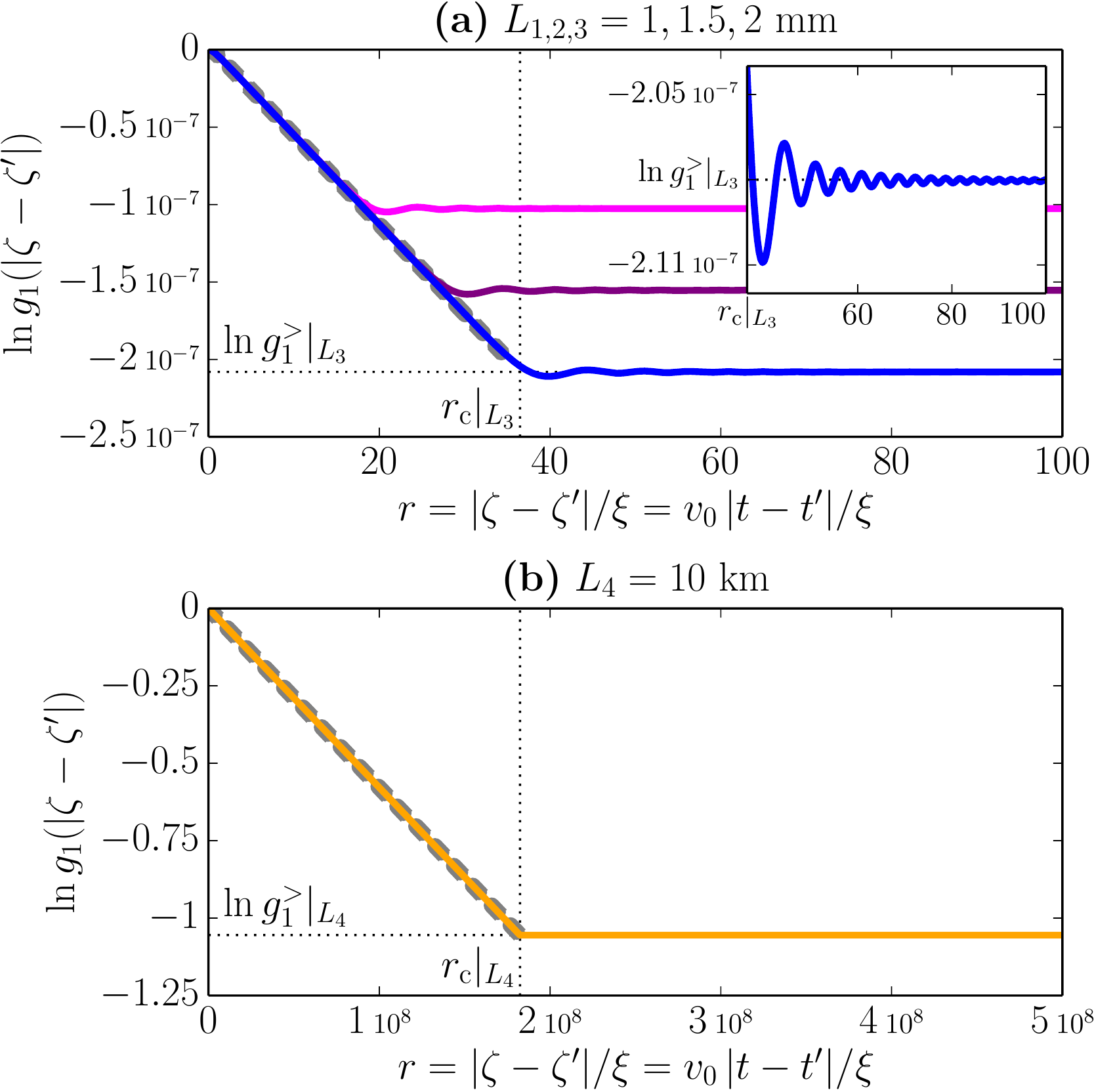}
\caption{Natural logarithm of the degree of first-order coherence $g_{1}(|\zeta-\zeta'|)$ of the light exiting the waveguide [Eq.~\eqref{Eq:DOFOCResultBis2}] as a function of the dimensionless relative distance $r$ [Eq.~\eqref{Eq:NormalizedRelativeDistance}]. The numerical parameters used for tracing the curves are listed in Tab.~\ref{Tab:ExperimentalParameters}. The magenta (purple, blue) curve in panel (a) and the orange one in panel (b) are obtained for $L=L_{1}=1~\mathrm{mm}$ ($L_{2}=1.5~\mathrm{mm}$, $L_{3}=2~\mathrm{mm}$) and $L_{4}=10~\mathrm{km}$ long, respectively, 1D nonlinear waveguides. The vertical (horizontal) dotted lines indicate the value of $r_{\mathrm{c}}$ defined by Eq.~\eqref{Eq:CriticalNormalizedRelativeDistance} [of $\ln g_{1}^{>}$ given by Eq.~\eqref{Eq:Aboverc2}] in the case where $L=L_{3}$: $r_{\mathrm{c}}|_{L_{3}}\simeq36.5$ ($\ln g_{1}^{>}|_{L_{3}}\simeq-2.1\times10^{-7}$), and in the case where $L=L_{4}$: $r_{\mathrm{c}}|_{L_{4}}\simeq1.8\times10^{8}$ ($\ln g_{1}^{>}|_{L_{4}}\simeq-1.1$). The thick gray dashed curves in panels (a) and (b) correspond to the $L$-independent $r\lesssim r_{\mathrm{c}}$ approximation \eqref{Eq:Belowrc2} of $g_{1}$ and the inset in panel (a) focuses on the oscillatory behavior of $\ln g_{1}(|\zeta-\zeta'|)$ around $\ln g_{1}^{>}$ for $L=L_{3}$.}
\label{Fig:DOFOC}
\end{figure}

Figure \ref{Fig:DOFOC} displays the natural logarithm of the $g_{1}$ function at the exit of the medium [Eq.~\eqref{Eq:DOFOCResultBis2}] as a function of the dimensionless distance $r$ [Eq.~\eqref{Eq:NormalizedRelativeDistance}] for $L=L_{1,2,3}=1,1.5,2~\mathrm{mm}$ [see Fig.~\ref{Fig:DOFOC}(a)] and $L_{4}=10~\mathrm{km}$ [see Fig.~\ref{Fig:DOFOC}(b)] long waveguides. At the $z=L$ face of the waveguide, that is, $\mathrm{\Delta}\tau=L/v_{0}<+\infty$ seconds of evolution after the quench of the fluid of light at $z=0$ (where, in passing, laser coherence is not yet affected by the presence of the interface: $g_{1}\to1$ as $\mathrm{\Delta}\tau\to0$), $g_{1}$ decays mostly exponentially, as predicted by Eq.~\eqref{Eq:InteractingThermalBehavior} and indicated by the gray dashed curves in Fig.~\ref{Fig:DOFOC}, up to $r_{\mathrm{c}}=|\zeta-\zeta'|_{\mathrm{c}}/\xi$, whose position corresponds to the vertical dotted lines in the figure. Notice that for a fixed effective mass $m$ and a fixed interaction parameter $g$, the characteristic length of the exponential decay in Eq.~\eqref{Eq:InteractingThermalBehavior} ends up not depending on the light power $\mathcal{P}$ [use Eqs.~\eqref{Eq:NumberDensity}--\eqref{Eq:EffectiveTemperature}]; of course, the higher the value of the two-photon interaction constant $g$, the faster the thermal-like decay \eqref{Eq:InteractingThermalBehavior} is. Afterwards, for $r>r_{\mathrm{c}}$, $g_{1}$ stays approximately locked to the constant value $g_{1}(|\zeta-\zeta'|_{\mathrm{c}})$; in fact, it weakly oscillates [see the inset of Fig.~\ref{Fig:DOFOC}(a)] around its, numerically almost equivalent, constant $r\to+\infty$ limit $g_{1}^{>}$ given by Eq.~\eqref{Eq:Aboverc2} and indicated by the horizontal dotted lines in Fig.~\ref{Fig:DOFOC}.

\paragraph{Relaxation of the photon gas towards a prethermalized state.---}

\begin{figure}[t!]
\includegraphics[width=\linewidth]{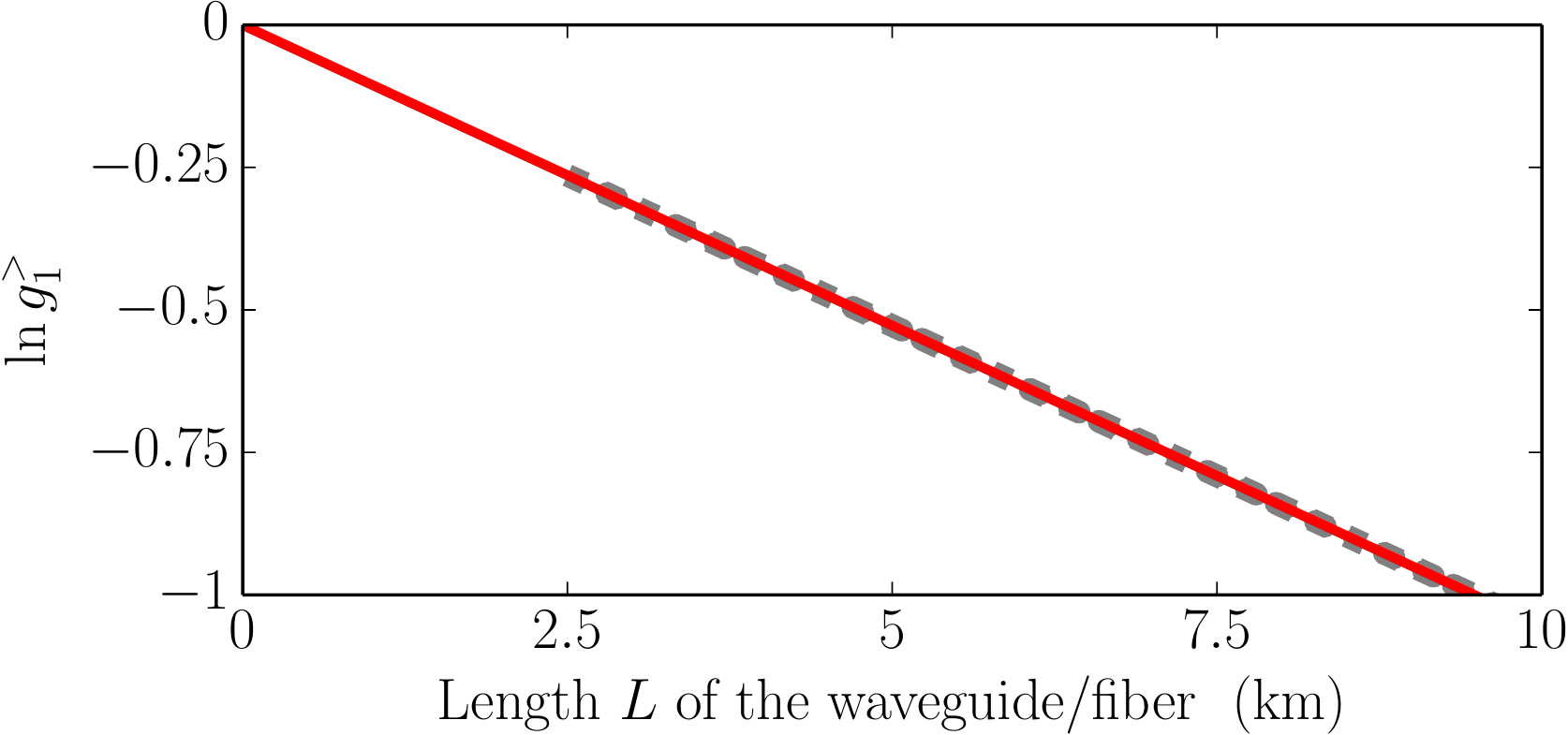}
\caption{Natural logarithm of the $r\to+\infty$ limit $g_{1}^{>}$ of the $g_{1}$ function, given by Eq.~\eqref{Eq:Aboverc2}, as a function of the waveguide's length $L$ (red curve). The thick gray dashed line corresponds to the large-$L$ behavior of $\ln g_{1}^{>}$, as predicted by Eq.~\eqref{Eq:LargeLAboverc}.}
\label{Fig:MerminWagner}
\end{figure}

As the time $\mathrm{\Delta}\tau=L/v_{0}$ after the quench of the quantum fluid of light at $z=0$ increases, that is, as longer and longer waveguides are considered, $r_{\mathrm{c}}$ increases too and the $r>r_{\mathrm{c}}$ plateau of the $g_{1}$ function decreases, which we shall discuss later with the help of Fig.~\ref{Fig:MerminWagner}. This evolution continues until the photon gas reaches, in the limiting case where $\mathrm{\Delta}\tau=+\infty$, a thermal-like state where $g_{1}$ is of the exponential form \eqref{Eq:InteractingThermalBehavior} across the entire 1D system, that is, as $\zeta$ is nothing but $t$ at a fixed propagation time $\tau$, over the entire duration of the measurement of the first-order coherence function. This prethermalized state corresponds to the relaxed, equilibrium, state of the 1D quantum fluid of light after the quench experienced at $z=0$. It emerges locally within the photon gas since the $r\lesssim r_{\mathrm{c}}$ exponential behavior of $g_{1}$ at different propagation times $\mathrm{\Delta}\tau<+\infty$ perfectly coincides with its exponential decay. According to Eq.~\eqref{Eq:EffectiveTemperature}, its effective temperature $T_{\mathrm{eff}}$ originates from the energy jump $|g|\,\rho=|g|\,\rho_{\mathrm{air}}/n_{0}$ at the $z=0$ air-waveguide interface. Using the realistic numerical parameters listed in Tab.~\ref{Tab:ExperimentalParameters}, one finds $T_{\mathrm{eff}}\simeq2.9~\mathrm{K}$.

\paragraph{Short-time Gaussian behavior of the coherence function of the transmitted light.---}

As predicted by the analytical formula \eqref{Eq:Belowrc2}, the whole $r\lesssim r_{\mathrm{c}}$ behavior of the coherence function $g_{1}$ at the exit of the waveguide (indicated by the gray dashed curves in Fig.~\ref{Fig:DOFOC}) is more complicated than the above-discussed thermal-like exponential decay \eqref{Eq:InteractingThermalBehavior}, the latter being in fact an approximation of Eq.~\eqref{Eq:Belowrc2} valid in the $1\ll r\lesssim r_{\mathrm{c}}$ regime. In the case where $r\ll1$, i.e., when $|\zeta-\zeta'|\ll\xi\Longleftrightarrow|t-t'|\ll\xi/v_{0}$, $g_{1}$ is Gaussian, given by Eq.~\eqref{Eq:NoninteractingThermalBehavior}. This behavior is typical of a noninteracting Bose system \cite{Pitaevskii2003, Naraschewski1999}, for which $g=0$ and, as a result, $\xi=+\infty$. However, within the range of numerical parameters given in Tab.~\ref{Tab:ExperimentalParameters} and as one may note from Fig.~\ref{Fig:DOFOC}, the $1\ll r\lesssim r_{\mathrm{c}}$ thermal-like exponential behavior \eqref{Eq:InteractingThermalBehavior} prevails upon the $r\ll1$ ``noninteracting'' Gaussian one \eqref{Eq:NoninteractingThermalBehavior}, the $r\ll 1$ limit corresponding to a tiny, even totally hidden in Fig.~\ref{Fig:DOFOC}(b), portion of the curves.

\paragraph{Light-cone-like spreading of the thermal correlations at the Bogoliubov sound velocity.---}

According to the two first paragraphs, a given point $\zeta$ in the 1D quantum fluid of light establishes thermal correlations with other points $\zeta'$ as long as the distance $|\zeta-\zeta'|$ is smaller than a certain characteristic value $|\zeta-\zeta'|_{\mathrm{c}}$. According to Eq.~\eqref{Eq:CriticalNormalizedRelativeDistance}, the latter linearly scales with the time $\mathrm{\Delta}\tau$ elapsed after the quantum quench as $|\zeta-\zeta'|_{\mathrm{c}}=2\,s\,\mathrm{\Delta}\tau$. This means that the prethermalized state of $g_{1}$ function \eqref{Eq:InteractingThermalBehavior} emerges in a light-cone-like evolution in the photon gas, the propagation velocity of the thermal correlations corresponding to the Bogoliubov speed of sound $s$ defined in Eq.~\eqref{Eq:SoundVelocity}. This may be understood as follows. As investigated in Ref.~\cite{Larre2015a}, the sudden modification of the optical nonlinearity at the entrance face of the dielectric is accompanied by a dynamical Casimir emission \cite{Moore1970, Fulling1976, Davies1977, Kardar1999} of pairs of correlated Bogoliubov excitations propagating in opposite directions along the $\zeta$ axis with wavenumbers $\pm\,k\gtrless0$, i.e. (cf.~explainations in the fifth paragraph of Sect.~\ref{SubSec:ModulusPhaseBogoliubovTheoryOfQuantumFluctuations}), of correlated paraxial photons whose angular frequencies $\omega_{0}\pm\delta\omega=\omega_{0}\mp v_{0}\,k\lessgtr\omega_{0}$ are symmetrically distributed around the laser pump's one $\omega_{0}$. As the modulus of their group velocity
\begin{equation}
\label{Eq:BogoliubovGroupVelocity}
v_{\mathrm{g}}(\pm\,k)=\frac{\partial\omega_{k}}{\partial k}(\pm\,k)=\pm\,s\,\frac{(\xi\,k)^{2}+2}{\sqrt{(\xi\,k)^{2}+4}}
\end{equation}
in the $\zeta\gtrless0$ direction is an increasing function of $k$, the Bogoliubov fluctuations forming each of the pairs emitted at $\tau=0$ establish thermal correlations at $\tau=\mathrm{\Delta}\tau$ as soon as their separation distance
\begin{equation}
\label{Eq:LightConeCondition}
|\zeta-\zeta'|<\underset{k}{\min}\,|v_{\mathrm{g}}(k)-v_{\mathrm{g}}(-k)|\,\mathrm{\Delta}\tau=2\,s\,\mathrm{\Delta}\tau,
\end{equation}
which corresponds to the previously-discussed light-cone condition. Considering the numerical parameters given in Tab.~\ref{Tab:ExperimentalParameters}, the Bogoliubov sound velocity $s\simeq7.0\times10^{5}~\mathrm{m}/\mathrm{s}$, which corresponds to a healing length $\xi\simeq9.2\times10^{-7}~\mathrm{m}$. For a $L=2~\mathrm{mm}$ long waveguide [blue curve in Fig.~\ref{Fig:DOFOC}(a)], one thus has $|\zeta-\zeta'|_{\mathrm{c}}\simeq3.4\times10^{-5}~\mathrm{m}$, and so $|t-t'|_{\mathrm{c}}=|\zeta-\zeta'|_{\mathrm{c}}/v_{0}\simeq4.0\times10^{-13}~\mathrm{s}$. In the case where $L=10~\mathrm{km}$ [Fig.~\ref{Fig:DOFOC}(b)], $|\zeta-\zeta'|_{\mathrm{c}}\simeq168.0~\mathrm{m}$ and $|t-t'|_{\mathrm{c}}\simeq2.0\times10^{-6}~\mathrm{s}$.

\paragraph{Loss of long-lived coherence in the course of the propagation along the waveguide.---}

When $|\zeta-\zeta'|>|\zeta-\zeta'|_{\mathrm{c}}$, that is, when $|t-t'|>|t-t'|_{\mathrm{c}}$, the coherence function $g_{1}$ stops decaying and is approximately locked to its $r\to+\infty$ limit $g_{1}^{>}$ given by the formula \eqref{Eq:Aboverc2} and indicated by the horizontal dotted lines in Figs.~\ref{Fig:DOFOC}(a, b) for $2~\mathrm{mm}$ and $10~\mathrm{km}$ long waveguides. Even though $g_{1}^{>}$ is nonzero, it is however smaller than unity: This indicates that the beam of light has partially lost its original long-lived coherence in the course of its propagation along the 1D nonlinear medium. This loss of coherence is expected to be significant for very long waveguides: Indeed, using Tab.~\ref{Tab:ExperimentalParameters}, $\ln g_{1}^{>}$ starts being smaller than $-1$ from a waveguide length $\sim L_{4}=10~\mathrm{km}$, for which one precisely has $\ln g_{1}^{>}\simeq-1.1$. This could have detrimental practical consequences, typically in fiber-optic communication where information has to be transmitted from one place to another over distances sometimes of the order of several hundreds of kilometers. In Fig.~\ref{Fig:MerminWagner},
we plot $\ln g_{1}^{>}$ against $L$. According to the asymptotic formula \eqref{Eq:LargeLAboverc}, the long-distance tail of the degree of first-order coherence is expected to exponentially tend to zero as $L$ increases:
\begin{equation}
\label{Eq:LargeLLongRangeBehavior}
g_{1}^{>}\underset{L\to+\infty}{\sim}\exp(-L/\ell_{\mathrm{c}}),
\end{equation}
with a coherence length
\begin{equation}
\label{Eq:CoherenceLength}
\ell_{\mathrm{c}}=\frac{c_{0}\,\varepsilon_{0}\,\rho_{\mathrm{air}}\,\xi^{2}}{\hbar\,\omega_{0}\,s}
\end{equation}
equal to $\simeq9.5\times10^{3}~\mathrm{m}$ given the numerical parameters of Tab.~\ref{Tab:ExperimentalParameters}. $g_{1}^{>}\to0$ when $L\to+\infty$, in full accordance with the Mermin--Wagner--Hohenberg--Coleman theorem \cite{Mermin1966, Hohenberg1967, Coleman1973} which stipulates that long-range order cannot exist in an infinite uniform 1D quantum fluid at nonzero temperature. According to Eq.~\eqref{Eq:CoherenceLength}, the coherence length $\ell_{\mathrm{c}}$ monotonically decreases with the light power $\mathcal{P}$ as $\ell_{\mathrm{c}}\propto\mathcal{P}^{-1/2}$; this is the direct consequence of the power-dependence of the healing length $\xi$ and of the Bogoliubov speed of sound $s$ [see Eqs.~\eqref{Eq:HealingLength} and \eqref{Eq:SoundVelocity}, respectively]. Given the wide range of waveguide lengths $L$ considered in Fig.~\ref{Fig:MerminWagner}, the small-$L$ behavior \eqref{Eq:LowLAboverc} of $g_{1}^{>}$, which should be visible for $L$'s much smaller than $v_{0}\,\xi/s\simeq1.1\times10^{-4}~\mathrm{m}$ (see the last paragraph of Sect.~\ref{SubSubSec:AnalyticalDerivation} and use Tab.~\ref{Tab:ExperimentalParameters}), cannot be displayed on the graph.

\paragraph{Oscillations around the plateau \eqref{Eq:Aboverc}.---}

As shown in the inset of Fig.~\ref{Fig:DOFOC}(a), the $r>r_{\mathrm{c}}$ part of the first-order coherence function displays weak-amplitude damped oscillations around the plateau $g_{1}^{>}$. Measuring on the graph the wavelength $\lambda_{\mathrm{osc}}$ of these oscillations, we find $\lambda_{\mathrm{osc}}\simeq4.8\,\xi$, which corresponds to a wavenumber $k_{\mathrm{osc}}=2\pi/\lambda_{\mathrm{osc}}\simeq1.3\,\xi^{-1}$, and so such that $\xi\,k_{\mathrm{osc}}\gtrsim1$. This indicates that the oscillation features displayed by $g_{1}$ originate from the contribution of the large-wavenumber, free-particle, Bogoliubov modes to the response of the fluid of light to the sudden quantum quench at $z=0$, as we are now going to explain. First of all, imagine that the switching on of the photon-photon interactions at the $z=0$ interface takes place on a finite length $\delta z$ along the $z$ axis, i.e., over a finite duration $\delta\tau=\delta z/v_{0}$, the sudden-quench case studied in this work being obviously recovered by taking the limit $\delta\tau\to0$. In the case where $\delta\tau\gg\hbar/(|g|\,\rho)$, the quench excites modes whose wavenumber $k$ (along the $\zeta=v_{0}\,t-z$ axis) belongs to the low-$k$ part of the Bogoliubov spectrum,
\begin{equation}
\label{Eq:MaxWavenumberFiniteDurationQuench}
|k|<k_{\ast}=\frac{1}{s\,\delta\tau}\ll\frac{1}{\xi},
\end{equation}
for which the dispersion law has the form $\omega_{k}\simeq s\,|k|$ (see Sect.~\ref{SubSec:ModulusPhaseBogoliubovTheoryOfQuantumFluctuations}), while all the other modes experience the finite-duration modulation of the optical nonlinearity as adiabatic \cite{Carusotto2010}. Correspondingly, the $g_{1}$ function \eqref{Eq:DOFOCResultBis1} may be approximated as
\begin{align}
\notag
&\left.g_{1}(|\zeta-\zeta'|)\simeq\exp\bigg\{{-}2\,\frac{v_{0}}{c_{0}}\,\frac{\hbar\,\omega_{0}}{\varepsilon_{0}\,\rho_{\mathrm{air}}}\,\Big(\frac{g\,\rho}{\hbar}\Big)^{2}\right. \\
\label{Eq:DOFOCFiniteDurationQuench}
&\left.\times\int\limits_{-k_{\ast}}^{k_{\ast}}\frac{\mathrm{d}k}{2\pi}\,\bigg[\frac{\sin(s\,|k|\,\mathrm{\Delta}\tau)}{s\,|k|}\bigg]^{2}\,[1-\cos(k\,|\zeta-\zeta'|)]\bigg\}.\right.
\end{align}
Taking the $k_{\ast}\to+\infty$ limit in this latter equation corresponds to performing a sudden quench [cf.~Eq.~\eqref{Eq:MaxWavenumberFiniteDurationQuench}] while sticking to a phononic approximation. In this case, the integral in Eq.~\eqref{Eq:DOFOCFiniteDurationQuench} is a two-step trapezoidal function of $|\zeta-\zeta'|$ which linearly decreases up to $|\zeta-\zeta'|=2\,s\,\mathrm{\Delta}\tau$ and which stays constant above, roughly as in Fig.~\ref{Fig:DOFOC}(a), but with an angular sliding edge at $|\zeta-\zeta'|=2\,s\,\mathrm{\Delta}\tau$ and without oscillations for $|\zeta-\zeta'|>2\,s\,\mathrm{\Delta}\tau$: A low-wavenumber, phononic, description of the response to the sudden quench we have at the entrance face of the waveguide predicts no oscillation feature in the $g_{1}$ function. As a consequence, the oscillations around the partial-coherence plateau $g_{1}^{>}$ that we observe in Fig.~\ref{Fig:DOFOC}(a) are due to the large-wavenumber, free-particle, Bogoliubov modes excited in the quenched quantum fluid of light.

\paragraph{Effect of photon absorption.---}

In the present work, we have focused on the case of a lossless fiber, for which the evolution of the light field is conservative. However, in an actual waveguide, photon absorption, and so dissipation, is well and truly present. In this final paragraph, we briefly establish the condition under which one can still apply the conservative evolution law \eqref{Eq:HarmonicEvolution} to describe the propagation of the quantum fluctuations of the beam of light along an actual, dissipative, fiber. For simplicity's sake, one will only consider linear losses, that is, one-photon absorption \cite{Agrawal1995}, which results in an exponential decay of the beam's power $\mathcal{P}=\frac{1}{2}\,c_{0}\,\varepsilon_{0}\,n_{0}\,\rho$ as a function of the propagation distance $0<z=v_{0}\,\tau<L=v_{0}\,\mathrm{\Delta}\tau$ along the waveguide:
\begin{equation}
\label{Eq:BackgroundDensityLosses}
\mathcal{P}(z)=\mathcal{P}_{0}\,\mathrm{e}^{-\alpha_{\mathrm{dB}}z/10},\quad\text{i.e.},\quad\rho(\tau)=\rho_{0}\,\mathrm{e}^{-\mathrm{\Gamma}\tau},
\end{equation}
where $\alpha_{\mathrm{dB}}=10\,\mathrm{\Gamma}/v_{0}>0$ is the attenuation constant of the fiber, expressed in $\mathrm{dB}/\mathrm{m}$ \cite{Agrawal1995}. Assuming the adiabatic approximation, for which the dissipation rate $\mathrm{\Gamma}$ is much smaller than any other frequency scale of the photon fluid, the evolution of the Bogoliubov operators $\hat{b}_{k}(\tau)$ may be modeled by a Heisenberg--Langevin equation \cite{Louisell1974, Gardiner2000} of the form \cite{Busch2014, Grisins2015}
\begin{equation}
\label{Eq:HeisenbergLangevinEquation}
\frac{\partial\hat{b}_{k}}{\partial\tau}=-\mathrm{i}\,\omega_{k}(\tau)\,\hat{b}_{k}-\frac{\mathrm{\Gamma}}{2}\,\hat{b}_{k}+u_{k}(\tau)\,\hat{\xi}_{k}^{\vphantom{\dag}}-v_{k}(\tau)\,\hat{\xi}_{-k}^{\dag}.
\end{equation}
In this equation,
\begin{equation}
\label{Eq:BogoliubovSpectrumLosses}
\hbar\,\omega_{k}(\tau)=\sqrt{\hbar\,\mathrm{\Omega}_{k}\,[\hbar\,\mathrm{\Omega}_{k}+2\,g\,\rho(\tau)]},
\end{equation}
and
\begin{equation}
\label{Eq:BogoliubovAmplitudesLosses}
u_{k}(\tau),v_{k}(\tau)=\frac{1}{2}\,\frac{\mathrm{\Omega}_{k}\pm\omega_{k}(\tau)}{\sqrt{\mathrm{\Omega}_{k}\,\omega_{k}(\tau)}}
\end{equation}
are the instantaneous Bogoliubov energy and Bogoliubov amplitudes, respectively, and $\hat{\xi}_{k}(\tau)$ is a quantum-noise operator which commutes with the $\hat{b}_{k}(\tau)$'s and which satisfies the commutation rules
\begin{subequations}
\label{Eq:CommutatorsNoise}
\begin{align}
\label{Eq:CommutatorsNoise1}
[\hat{\xi}_{k^{\vphantom{\prime}}}^{\vphantom{\dag}}(\tau),\hat{\xi}_{k'}^{\dag}(\tau')]&=2\pi\,\mathcal{N}\,\frac{\hbar\,\omega_{0}}{\varepsilon_{0}}\,\mathrm{\Gamma}\,\delta(k-k')\,\delta(\tau-\tau'), \\
\label{Eq:CommutatorsNoise2}
[\hat{\xi}_{k}(\tau),\hat{\xi}_{k'}(\tau')]&=0
\end{align}
\end{subequations}
at different times $\tau$, $\tau'$ to make the solution of Eq.~\eqref{Eq:HeisenbergLangevinEquation}, namely,
\begin{align}
\notag
\hat{b}_{k}(\tau)&\left.=\mathrm{e}^{-\mathrm{i}\int_{0}^{\tau}\mathrm{d}\tau'\,\omega_{k}(\tau')}\,\mathrm{e}^{-\mathrm{\Gamma}\tau/2}\,\hat{b}_{k}^{\mathrm{in}}\right. \\
\notag
&\left.\hphantom{=}+\int_{0}^{\tau}\mathrm{d}\tau'\,\mathrm{e}^{-\mathrm{i}\int_{\tau'}^{\tau}\mathrm{d}\tau''\,\omega_{k}(\tau'')}\,\mathrm{e}^{-\mathrm{\Gamma}(\tau-\tau')/2}\right. \\
\label{Eq:HeisenbergLangevinSolution}
&\left.\hphantom{=}\times[u_{k}(\tau')\,\hat{\xi}_{k}^{\vphantom{\dag}}(\tau')-v_{k}(\tau')\,\hat{\xi}_{-k}^{\dag}(\tau')],\right.
\end{align}
obey the algebra \eqref{Eq:CommutationRelationsElementaryExcitations} at all times $\tau$. According to Eqs.~\eqref{Eq:BackgroundDensityLosses}, \eqref{Eq:BogoliubovSpectrumLosses}, and \eqref{Eq:BogoliubovAmplitudesLosses}, Eq.~\eqref{Eq:HeisenbergLangevinSolution} may be approximated by the conservative, dissipationless, law \eqref{Eq:HarmonicEvolution} as soon as $\mathrm{\Gamma}\,\tau<1$ for all $\tau$, and so, \textit{a fortiori}, as soon as $\mathrm{\Gamma}\,\mathrm{\Delta}\tau<1$, that is, $\alpha_{\mathrm{dB}}\,L/10<1$: This imposes the maximum value $L_{\mathrm{max}}=10/\alpha_{\mathrm{dB}}$ of fiber length over which one can safely ignore the effect of photon absorption at the quantum-fluctuation level. While this condition is not straightforwardly satisfied for silicon-based devices, other material choices will provide the required level of transparency. The detailed investigation of the rich post-quench dissipative quantum dynamics in longer 1D waveguides will be the subject of future works.

\paragraph{Concluding remarks.---}

The results presented above extend to the present 1D paraxial-optics configuration recent experimental studies on the relaxation dynamics of a quenched phase-fluctuating ultracold 1D Bose gas \cite{Gring2012, Kuhnert2013, Langen2013}, as well as theoretical works on prethermalization in generic many-body 1D quantum systems \cite{Rigol2007}, in Fermi \cite{Moeckel2008, Moeckel2009, Moeckel2010, Marino2012} and Luttinger liquids \cite{Mitra2013, Buchhold2015}, in long-range quantum Ising models \cite{VanDenWorm2013, Marcuzzi2013}, on the light-cone-like spreading of two-point correlations following a quantum quench \cite{Calabrese2006, Bravyi2006, Calabrese2007, Lauchli2008, Carusotto2010, Mathey2010, Cheneau2012}, and on the links between this effect and the local emergence of thermal-like features \cite{Cramer2008}. In the present optical context, the quench of the system's Hamiltonian simply originates from the fact that the nonlinear dielectric possesses an entrance face\footnote{As discussed in the introduction of Sect.~\ref{Sec:LightCoherenceInResponseToQuantumQuenchesInTheKerrNonlinearity} and in Sect.~\ref{SubSec:PhysicalSituation}, the photon fluid is also quenched at the exit of the waveguide. This is fully accounted for in our analytical description of the $g_{1}$ function, but as we focus our attention on what happens at the very back face of the medium, just after the second quench, only the consideration of the first quantum quench at the $z=0$ entrance face matters for the present discussion.}; in that respect---as the quench protocol is provided by the very nature of the used platform---no extra manipulation on the system needs to be implemented by the experimentalist to get the prethermalization features shown in Fig.~\ref{Fig:DOFOC}. This illustrates the power of nonlinear-waveguide setups as a simple platform to study quantum dynamical aspects in quenched many-body 1D Bose systems. Finally, it is of paramount importance to test the validity of the two main hypothesis of the present study---namely, the single-transverse-mode condition \eqref{Eq:SingleTransverseModeConditionExperiments} and the dilute-gas constraint \eqref{Eq:DiluteGasConditionExperiments}---for the realistic optical parameters given in Tab.~\ref{Tab:ExperimentalParameters}. On the one hand, one checks that the right-hand side of the inequality \eqref{Eq:SingleTransverseModeConditionExperiments} is approximately $10^{2}$ times larger than the nonlinearity parameter $|\mathrm{\Delta}n_{\mathrm{NL}}|$, validating the single-transverse-mode approximation \eqref{Eq:SingleTransverseModeConditionExperiments} and so the classical (quantum) 1D Gross--Pitaevskii equation \eqref{Eq:ClassicalGPE} [\eqref{Eq:QuantumGPE}]. On the other hand, the left-hand side of \eqref{Eq:DiluteGasConditionExperiments} is around $10^{15}$ times smaller than $|\mathrm{\Delta}n_{\mathrm{NL}}|$, which ensures that the considered photon fluid is (very) dilute\footnote{Given the numerical values of Tab.~\ref{Tab:ExperimentalParameters}, one may show that the inequality \eqref{Eq:DiluteGasConditionExperiments} starts being no longer valid from input optical powers $\mathcal{P}\lesssim1~\mathrm{pW}$. In that case, under suitable conditions on the incident beam's coherence and on the injection mechanism \cite{LebreuillyToBePublished, LebreuillyBisToBePublished}, the 1D photon fluid is expected to enter the strongly interacting, Tonks--Girardeau, regime.} and, as a result, that its quantum fluctuations can be described with a good accuracy by means of the modulus-phase Bogoliubov theory of weakly interacting Bose gases reviewed in Sect.~\ref{SubSec:ModulusPhaseBogoliubovTheoryOfQuantumFluctuations}.

\section{Conclusion}
\label{Sec:Conclusion}

Using a general quantum theory of paraxial light propagation in 1D nonlinear geometries, we have investigated the coherence properties of a beam of light emerging from a finite-length single-mode 1D waveguide presenting a weak Kerr nonlinearity. By mapping the longitudinal propagation of the optical field in the increasing-$z$ direction onto a time evolution and by identifying the actual time parameter $t$ as a spatial coordinate, we have entirely reformulated our predictions in the language of many-body physics, as the response of a 1D fluid of many weakly interacting photons to a pair of quantum quenches in the photon-photon interaction constant, at the entrance and the exit of the nonlinear waveguide. Pursuing along the lines of Ref.~\cite{Larre2015a}, this makes it possible to illustrate the potential of nonlinear propagating geometries as novel, simple, platforms for studying quantum dynamical features in quenched many-body Bose systems.

At the exit of the waveguide, the degree of first-order coherence of the transmitted light features the occurrence of a relaxation dynamics of the quantum fluid of light towards a prethermalized state typical of quenched quantum systems of weakly interacting bosons \cite{Gring2012, Kuhnert2013, Langen2013, Rigol2007, Moeckel2008, Moeckel2009, Moeckel2010, Marino2012, Mitra2013, Buchhold2015, VanDenWorm2013, Marcuzzi2013, Calabrese2006, Bravyi2006, Calabrese2007, Lauchli2008, Carusotto2010, Mathey2010, Cheneau2012, Cramer2008}. The corresponding thermal correlations emerge locally in their final form in the photon gas, spreading in a light-cone-like evolution at the Bogoliubov speed of sound. This directly results in a loss of long-lived coherence in the transmitted beam of light, which could have detrimental practical consequences in telecommunication \textit{via} kilometer-long optical fibers.

While the present study mainly focuses on the weak-interaction regime, the general 1D quantum theory actually holds for any values of the 1D two-photon interaction constant. As a result, it may be used for studying quantum dynamical features in the strongly interacting, Tonks--Girardeau, regime, which may be obtained for both large 1D Kerr-nonlinearity coefficients and low optical powers. Its theoretical investigation in a typical nonlinear-optics configuration will be the subject of forthcoming publications \cite{LebreuillyToBePublished, LebreuillyBisToBePublished}.

\begin{acknowledgement}

We acknowledge Alessio Chiocchetta, Pjotrs Gri\v{s}ins, and Jos\'e Lebreuilly for our continuous exchanges on the physics of quantum quenches, Stefano Biasi and Fernando R.~Manzano for our stimulating discussions on the possibility of an experimental realization, Nicolas Pavloff for his careful reading of the paper and his helpful comments, and Andrea Gambassi for having indicated us relevant references about prethermalization. This work was supported by the ERC through the QGBE grant, by the EU-FET Proactive grant AQuS, Project No.~640800, and by the Autonomous Province of Trento, partially through the SiQuro project (``On silicon chip quantum optics for quantum computing and secure communications'').

\end{acknowledgement}

\section*{Appendix: Analytical results for optical potentials $U(\mathbf{x})$ of experimental interest}

\subsection*{Parabolic potential}

An optical confinement of parabolic shape may be realized when
\begin{subequations}
\label{Eq:HarmonicConfinement}
\begin{align}
\label{Eq:HarmonicConfinement1}
\mathrm{\Delta}n(\mathbf{x},\omega)&=-\tfrac{1}{2}\,n(\omega)\,\varkappa^{2}\,\mathbf{x}^{2}, \\
\label{Eq:HarmonicConfinement2}
U(\mathbf{x})&=\tfrac{1}{2}\,\beta_{0}\,\varkappa^{2}\,\mathbf{x}^{2}~\text{[from the 1st of Eqs.~\eqref{Eq:ExternalPotential3DInteractionConstant}]},
\end{align}
\end{subequations}
where $\varkappa$ is a positive parameter controlling the strength of the confinement. By analogy with quantum-gas physics, this amounts to consider the well-known problem of atoms of mass $M$ confined along the $z$ axis by means of a transverse harmonic trapping potential $U_{\perp}(\mathbf{x})=\frac{1}{2}\,M\,\omega_{\perp}^{2}\,\mathbf{x}^{2}$ of oscillation frequency $\omega_{\perp}$. Solving the 2D Schr\"odinger-type equation $\kappa_{0}\,\mathrm{\Phi}_{0}=[-\nabla^{2}/(2\,\beta_{0})+U(\mathbf{x})]\,\mathrm{\Phi}_{0}$, one finds
\begin{equation}
\label{Eq:GroundStateHarmonicConfinement}
\kappa_{0}=\varkappa\quad\text{and}\quad\mathrm{\Phi}_{0}(\mathbf{x})=\sqrt{\frac{2}{\pi\,w^{2}}}\,\mathrm{e}^{-\mathbf{x}^{2}/w^{2}},
\end{equation}
where $w=(\beta_{0}\,\varkappa/2)^{-1/2}$. Correspondingly,
\begin{equation}
\label{Eq:EffectiveModeAreaHarmonicConfinement}
A_{\mathrm{eff}}=\pi\,w^{2}
\end{equation}
[cf.~Eq.~\eqref{Eq:EffectiveModeArea}], $E_{0}=\hbar\,v_{0}\,\varkappa$ [cf.~the definition of $E_{0}$ before Eq.~\eqref{Eq:InteractionConstant}], and, by means of Eqs.~\eqref{Eq:ExternalPotential3DInteractionConstant}, \eqref{Eq:1DInteractionConstant}, \eqref{Eq:InteractionConstant}, and \eqref{Eq:EffectiveModeAreaHarmonicConfinement},
\begin{equation}
\label{Eq:InteractionConstantHarmonicConfinement}
g=-\frac{\hbar\,k_{0}\,v_{0}\,n_{2}(\omega_{0})}{\pi\,w^{2}}.
\end{equation}
Furthermore, the dilute [constraint \eqref{Eq:DiluteGasConditionExperiments}] single-transverse-mode [constraint \eqref{Eq:SingleTransverseModeConditionExperiments}] regime is obtained when
\begin{equation}
\label{Eq:SuperfluidDiluteRegimeHarmonicConfinement}
\frac{2\,(\hbar\,\omega_{0})^{2}\,k_{0}\,|\tilde{n}_{2}(\omega_{0})|^{2}}{\pi^{2}\,|D_{0}|\,w^{4}}\overset{\eqref{Eq:DiluteGasConditionExperiments}}{\ll}|\mathrm{\Delta}n_{\mathrm{NL}}|\overset{\eqref{Eq:SingleTransverseModeConditionExperiments}}{\ll}\frac{2}{n_{0}^{\vphantom{2}}\,k_{0}^{2}\,w^{2}}.
\end{equation}

\subsection*{Square-well potential}

In the case where the waveguide is composed of a core of squared transverse cross section $a\times a$ and refractive index $n(\omega)$ transversally surrounded by a cladding of refractive index $N(\omega)<n(\omega)$, one has
\begin{subequations}
\label{Eq:SquareWellConfinement}
\begin{align}
\label{Eq:SquareWellConfinement1}
\mathrm{\Delta}n(\mathbf{x},\omega)&=
\begin{cases}
0                   & (|x|,|y|\leqslant a/2) \\
N(\omega)-n(\omega) & (|x|,|y|>a/2)
\end{cases}
, \\
\label{Eq:SquareWellConfinement2}
U(\mathbf{x})&=
\begin{cases}
0     & (|x|,|y|\leqslant a/2) \\
U_{0} & (|x|,|y|>a/2)
\end{cases}
,
\end{align}
\end{subequations}
where, using the first of Eqs.~\eqref{Eq:ExternalPotential3DInteractionConstant}, $U_{0}=-k_{0}\,(N_{0}-n_{0})>0$ [with $N_{0}=N(\omega_{0})$] and we have assumed that the cladding is approximately infinite in the transverse $x$ and $y$ directions. This amounts to consider the well-known problem of a quantum particle of wavefunction $\mathrm{\Phi}_{0}(\mathbf{x})$ trapped in a 2D potential well of barrier energy $0<U_{0}\leqslant+\infty$. In order to facilitate the analytical resolution of the Schr\"odinger-type equation $\kappa_{0}\,\mathrm{\Phi}_{0}=[-\nabla^{2}/(2\,\beta_{0})+U(\mathbf{x})]\,\mathrm{\Phi}_{0}$ supplemented by Eq.~\eqref{Eq:SquareWellConfinement2}, one takes the limit $U_{0}\to+\infty$. In this case, one finds the well-known results
\begin{subequations}
\label{Eq:GroundStateSquareWellConfinement}
\begin{align}
\label{Eq:GroundStateSquareWellConfinement1}
\kappa_{0}&=\frac{\pi^{2}}{\beta_{0}\,a^{2}}\quad\text{and} \\
\label{Eq:GroundStateSquareWellConfinement2}
\mathrm{\Phi}_{0}(\mathbf{x})&=
\begin{cases}
\displaystyle{\frac{2}{a}\cos\Big(\frac{\pi\,x}{a}\Big)\cos\Big(\frac{\pi\,y}{a}\Big)} & (|x|,|y|\leqslant a/2)
\vspace{1mm} \\
0                                                                                      & (|x|,|y|>a/2)
\end{cases}
,
\end{align}
\end{subequations}
yielding
\begin{equation}
\label{Eq:EffectiveModeAreaSquareWellConfinement}
A_{\mathrm{eff}}=\frac{4}{9}\,a^{2},
\end{equation}
$E_{0}=\pi^{2}\,\hbar\,v_{0}/(\beta_{0}\,a^{2})$, and
\begin{equation}
\label{Eq:InteractionConstantSquareWellConfinement}
g=-\frac{9\,\hbar\,k_{0}\,v_{0}\,n_{2}(\omega_{0})}{4\,a^{2}}.
\end{equation}
Within the dilute single-transverse-mode regime, the nonlinear shift $\mathrm{\Delta}n_{\mathrm{NL}}$ of the waveguide's refractive index defined in Eq.~\eqref{Eq:NonlinearRefractiveIndex} satisfies
\begin{equation}
\label{Eq:SuperfluidDiluteRegimeSquareWellConfinement}
\frac{81\,(\hbar\,\omega_{0})^{2}\,k_{0}\,|\tilde{n}_{2}(\omega_{0})|^{2}}{8\,|D_{0}|\,a^{4}}\overset{\eqref{Eq:DiluteGasConditionExperiments}}{\ll}|\mathrm{\Delta}n_{\mathrm{NL}}|\overset{\eqref{Eq:SingleTransverseModeConditionExperiments}}{\ll}\frac{\pi^{2}}{n_{0}^{\vphantom{2}}\,k_{0}^{2}\,a^{2}}.
\end{equation}
When $\kappa_{0}<U_{0}<+\infty$, the modal function $\mathrm{\Phi}_{0}(\mathbf{x})$ is approximately of the form \eqref{Eq:GroundStateSquareWellConfinement2} in the core of the optical waveguide and it exponentially decreases to zero outside.

\end{document}